\documentclass[%
preprint,
nofootinbib,
 amsmath,amssymb,
]{revtex4-2}

\usepackage{graphicx}
\usepackage{dcolumn}
\usepackage{bm}
\usepackage[mathlines]{lineno}

\newcommand{\OO}{\hat{\mathcal{O}}}

\newcommand{\ex}[1]{\langle #1\rangle}

\newcommand{\RR}{\mathbb{R}}
\newcommand{\CC}{\mathbb{C}}
\newcommand{\A}{\mathcal{A}}
\newcommand{\B}{\mathcal{B}}
\newcommand{\HH}{\mathcal{H}}
\newcommand{\ket}[1]{|#1\rangle}

\newcommand{\tr}{\operatorname{tr}}
\newcommand{\minprob}[3]{\begin{array}{ll}
	\operatorname{min}_{#1} & #2 \\
	\text { s.t. } & #3 \\
\end{array}}

\usepackage{xcolor}

\begin{document}


\title{Numerical exploration of the bootstrap in spin chain models}

\author{David Berenstein}
 \email{dberens@physics.ucsb.edu}
 \affiliation{Department of Physics, UC Santa Barbara 93106}
 \affiliation{Institute of Physics, University of Amsterdam, Science Park 904, PO Box 94485, 1090 GL Amsterdam,
The Netherlands}
\affiliation{Delta Institute for Theoretical Physics, Science Park 904, PO Box 94485, 1090 GL Amsterdam, The
Netherlands}
\author{George Hulsey}%
 \email{hulsey@physics.ucsb.edu}
\affiliation{Department of Physics, UC Santa Barbara 93106}%

\author{P. N. Thomas Lloyd}%
 \email{plloyd@ucsb.edu}
\affiliation{Department of Physics, UC Santa Barbara 93106}%

\date{\today}

\begin{abstract}
We analyze the bootstrap approach (a dual optimization method to the variational approach) to one-dimensional spin chains, leveraging semidefinite programming to extract numerical results. We study how correlation functions in the ground state converge to their true values at and away from criticality and at relaxed optimality. We consider the transverse Ising model, the three state Potts model, and other non-integrable spin chains  and investigate to what extent semidefinite methods can reliably extract numerical emergent physical data, including conformal central charges, correlation lengths and scaling dimensions. We demonstrate procedures to extract these data and show preliminary results in the various models considered. We compare to exact analytical results and to exact diagonalization when the system volume is small enough.
When we attempt to go to the thermodynamic limit, the semidefinite numerical method with translation invariance imposed as a constraint finds the solution with periodic boundary conditions even if these have not been specified. This implies that the determination of all conformal data in correlators has to be handled at finite volume.
Our investigation reveals that the approach has practical challenges. In particular, the correlation functions extracted from the optimal solution, which function as slack variables in the optimization, have convergence issues that suggest an underlying exponential complexity in the system size.

\end{abstract}

\maketitle



\section{Introduction}
The bootstrap program takes a constraint-oriented view of quantum mechanics. Its success comes from using very general quantum mechanical principles---unitarity, symmetry, algebraic structure---to make rigorous numerical or analytical statements about observables in quantum theories. Historically, the approach had its modern inception in conformal field theory with the development of the analytical, and later, numerical conformal bootstrap \cite{Belavin_Polyakov_Zamolodchikov_1984,Alday_Zhiboedov_2017,Kos_Poland_Simmons-Duffin_2014,Simmons-Duffin_2015,PhysRevD.86.025022}. The strict requirements of conformal symmetry lead to a tower of constraints on correlation functions which, when combined with unitarity, allows one to determine conformal data (scaling dimensions, operator-product expansion coefficients) with increasing precision. 

However, there are many systems without conformal symmetry which still enjoy sufficient quantum mechanical structure to be amenable to a bootstrap approach. In recent years, the scope of bootstrap approaches has widened to include matrix models, matrix quantum mechanics, lattice Yang-Mills theory, Schrodinger quantum mechanics, and more \cite{Aikawa:2021eai,Berenstein:2021dyf,Berenstein:2022unr,Berenstein:2022ygg,Bhattacharya:2021btd,Han:2020bkb,Kazakov:2022xuh,kazakovMM,Hu:2022keu,Tchoumakov:2021mnh,Lin:2020mme,Lawrence:2021msm}. In this melange of systems the bootstrap represents a very general approach to quantum mechanics which does not rely on standard notions of integrability or perturbative control; the absence of such properties generically hinders the possibility of having analytical results. Instead, the bootstrap takes advantage of the algebraic structure of the problem, the details of which are usually known \textit{a priori} and the manipulations of which are easily encodable on a computer. When coupled with polynomially efficient modern optimization algorithms, such as the primal-dual interior point solvers used in the conformal bootstrap \cite{Simmons-Duffin_2015}, bootstrap methods offer a rigorous numerical approach which challenges the state of the art in systems where traditional numerical methods--e.g. Monte Carlo--may be inefficient, inexact, or complicated by the ``sign problem" \cite{Troyer_Wiese_2005}. 

In this work we bring the numerical bootstrap program full circle by attempting to use it to extract numerical conformal data from critical quantum spin chains. Unlike the conformal bootstrap, we do this by studying finite-size spin chains at criticality and regressing their properties---energy and correlation functions---with the known, universal finite-size behavior. This work is directly inspired by the preliminary investigations along these lines carried out in \cite{Lawrence:2021msm}, which extends the bootstrap program in quantum mechanics and matrix models to many-body systems and subsequent work on similar methods on bootstrap bounds on the gap by the group of \cite{Nancarrow:2022wdr}. 

The main idea is to minimize the energy given a subset (a relaxation) of all the positivity constraints of a quantum system. The positivity constraint is that the expectation value of a positive operator is positive. If one considers all positive operators, one solves all constraints. A relaxation of the constraints is when we only impose some of the positive constraints, but not all.
The minimum of the energy subject to the positive constraints is an approximation of the ground state energy of the system.
Since the true ground state satisfies all possible positivity constraints, including the subset of relaxed constraints, the method provides a numerically rigorous lower bound for the ground state energy. The variables that enter the optimization problem are actually the correlation functions themselves. The solution of the optimization problem therefore has some information about the physics of the ground state beyond the ground state energy that one is trying to optimize. The question is then the convergence of these variables, the putative correlation functions, to the corresponding correlators in the actual ground state. 

Within this paradigm, we construct hierarchies of non-commutative optimization problems and solve them to determine conformal weights and central charges for two integrable systems where the conformal data is exactly known: the transverse-field Ising model and the three-state Potts model. With the correct setup, albeit with some difficulty, the approach can yield reasonably good numerical results in these well-understood models.  We then apply these same approaches to a non-integrable model that deforms the transverse Ising problem at criticality to see how well the approach generalizes. 

Our investigations reveal that there exist significant challenges to reliably estimating critical exponents and the central charge in generic spin systems. Moreover, the convergence of these methods is  not  well understood. 
We compare our bootstrap approach with other semidefinite methods from the condensed matter literature, where it is known that certain aspects of the problem fall in hard quantum complexity classes. Other methods for the extraction of conformal data, such as the lattice modes and periodic uniform matrix-product states approach of \cite{Zou_2018,Zou_2020,Milsted_Vidal_2017}, at present provide a more robust solution to determining CFT spectra from critical spin systems. However, the approach taken by these works relies on an exact or approximate diagonalization of the full Hamiltonian. The power of the semidefinite approach is its ability to study only correlations of operators in the state, while remaining agnostic to the state itself. Another important benefit of the bootstrap is that it provides mathematically rigorous bounds on the energy of the system, and some of its generalizations \cite{Fawzi_Fawzi_Scalet_2023b}. Most of the times these rigorous bounds are too loose to be able to do precision measurements.

The paper is organized as follows. In section \ref{sec:boot} we review the methods we will use and provide a description of  related works. In section \ref{sec:TIM} we discuss our implementation of the transverse Ising model using these methods. Because the system can be solved exactly, we  compare the results of the numerical method and the exact answer for various correlation functions. We  compare to the conformal data and other physical quantities from the numerical date we generated. We also discuss how consequences of the non-invertible symmetry at criticality are seen by the numerical method. We find that the system naturally wants to be described in terms of  fermion variables. In section \ref{sec:pots} we study the three state Pots model. Although the system at criticality is integrable, it is not free. The analog of the free fermion variables (the Fradkin-Kadanof parafermions) fail to solve the problem exactly. We find reasonable qualitative agreement, but so far we do not have a good determination of the critical exponents. In section \ref{sec:ANNI} we study a deformation of the Ising model that preserves the non-invertible symmetry but breaks integrability, the ANNNI model. We show that the free fermion basis works in a small vicinity of the Ising model, but the results deteriorate quickly as we break integrability further. We conclude in \ref{sec:Con}.

\section{The bootstrap and other semidefinite methods} \label{sec:boot}
Determining the ground state of a quantum mechanical system is a canonical example of a difficult but convex optimization problem. Usually, `solving' a quantum system is understood to be the problem of determining a state $\ket{\psi}$ in a physical Hilbert space $\HH$ subject to the constraints of unitarity, $\langle \psi | \psi \rangle = 1$, and any dynamical relations furnished by the Hamiltonian $\hat{H}$. One understands operators as mapping states to other states. 
However, one may also view the problem algebraically. The data of the generic quantum mechanical problem may be restated as an algebra of observables $\A$ (which act on the Hilbert space) and a set of linear functionals $\omega_\psi:\A \to \RR$ which represent the physical states. Here, one views states as acting on operators.

Formally, the bootstrap and variational approaches represent dual simplifications of the same fundamental, exact optimization problem. To make this precise, consider an interacting quantum system with many degrees of freedom. Any physical state defines, or is equivalently defined by, a Hermitian positive semidefinite (PSD) density operator $\rho \succeq 0$. The ground state of a system with Hamiltonian $\hat{H}$ is the optimum $E^\star$ of the convex optimization problem
\begin{equation}
    \label{eq:fundamentalOpt}
    \minprob{\rho}{\tr(\rho \hat{H}),}{\rho \succeq 0;\ \tr(\rho) = 1,}
\end{equation}
where the optimization occurs over the positive semidefinite  (PSD) cone of all physical density matrices, defining a semidefinite program (there is additionally a Hermiticity constraint $\rho^\dagger = \rho$ which we take to be implicit in the semidefinite constraint throughout) \cite{boyd}. While formally simple, this optimization problem generically contains an exponential number of degrees of freedom in the physical size of the system. For example, if the system is composed of $N$ qubits, a generic density matrix is specified by $\mathcal{O}(2^{2N})$ parameters all coupled by the semidefinite constraint, making a direct solution of \eqref{eq:fundamentalOpt} infeasible for all but the smallest systems.  

In order to make the program computationally feasible, one must somehow simplify the semidefinite constraint. Broadly speaking, there are two ways to do this: by either restricting or enlarging the feasible domain (recall that the feasible domain is the set of optimization variables where the constraints are satisfied). The variational approach restricts the feasible domain and views the problem in terms of states. Any variational ansatz---matrix product states, projected entangled-pair states---is a parametrization of a set of density matrices $\rho_\theta$ by a linear or at worst polynomial (in the system size $N$) set of parameters $\theta$. These states are usually constructed to be PSD and normalizable and hence the problem \eqref{eq:fundamentalOpt} becomes the much simpler optimization problem
\begin{equation}
    \label{eq:variationalOpt}
    \minprob{\theta}{\tr(\rho_\theta \hat{H}),}{\rho_\theta \succeq 0;\ \tr(\rho_\theta) = 1,}
\end{equation}
over the variational parameters. When we say simpler, we are referring to the dimension of the optimization domain---the parametrization may make the optimization over $\theta$ non-convex and non-linear, in which case different optimization methods like stochastic gradient descent or simulated annealing are needed to solve the problem. A price of this parametrization is that the optimum $E_\theta^\star$ of the program \eqref{eq:variationalOpt} gives only an upper bound on the true ground state energy: $E_\theta^\star \geq E^\star$. This is because any sub-exponential parametrization of states $\rho_\theta$ necessarily covers only a strict subspace of the entire physical Hilbert space $\HH_\theta \subset \HH$. Good ansatzes will offer a sharp upper bound, but the feasible domain will (almost) never extend to cover the exact ground state. 

While the variational approach restricts the feasible domain to the space of variational parameters, bootstrap methods instead \textit{relax} the semidefinite constraint by enlarging the feasible domain. Informally, this means enforcing positivity only on some submatrix of the entire density matrix $\rho$. This can be realized as follows. The semidefinite constraint $\rho \succeq 0$ is equivalent to the positivity of $\tr(\rho \OO) \geq 0$ for all positive semidefinite operators $\OO\succeq 0$ in the observables algebra $\A$, a subset we can call $\A_{+}$. To relax the constraint, one enforces this positivity condition on only a subset of such operators $\mathcal{S} \subset \A_+$, defining the semidefinite program
\begin{equation}
    \label{eq:relaxedOpt}
    \minprob{\rho}{\tr(\rho \hat{H}),}{\tr(\rho\mathcal{S}_i)\geq 0;\ \tr(\rho) = 1.}
\end{equation}
Whereas in the variational approach we optimized over a subspace of the physical Hilbert space, our relaxed constraint now means we are optimizing over a larger Hilbert space which contains possibly unphysical states: $\{\tr(\rho\mathcal{S})\geq 0\} \supset \{\tr(\rho \A) \geq0 \}$. Therefore, the optimum $E_\mathcal{S}^\star$ of the program \eqref{eq:relaxedOpt} always give a lower bound on the true optimum: $E_\mathcal{S}^\star \leq E^\star$. 

The bootstrap and the variational principle are in this sense dual approximations of the same fundamental optimization problem. While the challenge in the variational approach is the selection of a ``good" variational ansatz (a subset of states), the challenge in the bootstrap approach is the selection of a ``good" subset of operators $\mathcal{S}$ on which to enforce the positivity constraint.
\subsection{The many-body bootstrap}
With the general motivation in mind, let us formalize the setup of interest. We consider a finite quantum spin chain on $N$ sites. In principle we can consider any number of dimensions but we confine ourselves to one spatial dimension throughout this work. To each site $n$ is associated a finite set of operators $\{\OO^{(i)}_n\}$ which collectively form an algebra $\A$ over the entire chain. In the case of a spin-1/2 chain, one has the Pauli matrices $I,X,Y,Z$ at each site, and the algebra is ${\mathbb R}^{\otimes 4N}$ as usual Let $\text{span}(\B) \subset \A$ be some subspace which contains the Hamiltonian; regarding the basis $\B$ as a vector of operators one may write
\begin{equation}
    \ex{\hat{H}} = c^T\ex{\B}
\end{equation}
for some (complex) vector $c$. Here, the expectation value should be viewed as a linear functional $\ex{\cdot}:\A \to \CC$ on the operator algebra; the space of such maps is the domain of the optimization. 

For any set of operators $\B$, we construct the moment matrix $M$, a $|\B|\times|\B|$ Hermitian matrix with elements
\begin{equation}
    M_{ij}(\B) = \ex{\B^\dagger_i\B_j}.
\end{equation}
The relaxed density matrix constraint of \eqref{eq:relaxedOpt} is equivalent to a matrix positivity condition on $M(\B)$:
\begin{equation}
    M(\B) \succeq 0. 
\end{equation}
This positivity condition is the same as that which appears in the classical moment problem \cite{Akhiezer_2020} and in most if not all of the previous literature on the bootstrap program outside of CFT. 

There may be other physical constraints. For example, if we wish to consider only energy eigenstates (which includes the ground state), one has the linear constraints
\begin{equation}
    \ex{[\hat{H},\OO]} = 0.
\end{equation}
for any operator $\OO$. Our earlier work on the bootstrap in Schrodinger quantum mechanics uncovered additional anomalous constraints, linear and semidefinite, which encode boundary conditions or other limiting information about the states under consideration \cite{Berenstein:2022ygg}. These constraints are all standard for bootstrap approaches and their importance to the output of the program depends on the specific physical system in question. Other possible additions may include constraints on the entropy or variance of states and observables \cite{Fawzi_Fawzi_Scalet_2023a,Haim_Kueng_Refael_2020}.

The bootstrap problem for the ground state energy is therefore the following semidefinite program: given an operator basis $\B$ and Hamiltonian $\hat{H}$, solve
\begin{equation}
    \label{eq:btsp_energy_prob}
    \minprob{}{\ex{\hat{H}} = c^T\ex{\B},}{M(\B)\succeq 0;\ \ex{[\hat{H},\B_i]} = 0.}
\end{equation}
A key idea of the approach is that of hierarchies. By considering increasingly large operator bases $\B$, the result of the optimization problem \eqref{eq:btsp_energy_prob} approaches the exact value, coinciding when $\B \cong \A$ or when certain conditions on the rank of the optimal $M$ are met \cite{Araujo_Klep_Vertesi_Garner_Navascues_2023}. Formally, for a flag of subspaces
\begin{equation}
    \{1\} \subset \text{span}(\B_1) \subset \text{span}(\B_2) \subset \cdots \subset \A,
\end{equation}
one has $E^\star_{\B_1} \leq E^\star_{\B_2} \leq \cdots \leq E^\star$. The semidefinite constraint at level 2 contains that at level 1 and so on. This hierarchy allows one to construct increasingly large programs the optima of which grow increasingly close to the correct value of the objective function \cite{Pironio_Navascues_Acin_2010}. If the bases in the hierarchy are chosen correctly, one may even be guaranteed exponential convergence toward the exact value in the size of the basis chosen \cite{Fawzi_Fawzi_Scalet_2023b}. In principle the correlation functions---the matrix elements of $M(\B)$---should also approach their exact value, but their convergence properties are not as well understood and experimentally appear weaker than those of the objective function. 

\subsection{Related work}
Semidefinite methods for many-body systems have been recognized and applied in the mathematical, condensed matter/quantum information literature. In this section, we review related work and approaches taken to this problem which come from outside the bootstrap program in high-energy theory. 
\subsubsection{Mathematical optimization}
From a mathematical perspective, the problem \eqref{eq:btsp_energy_prob} is one of non-commutative polynomial optimization. The general problem of polynomial optimization in commuting variables, i.e.
\begin{equation}
    \label{eq:polyopt}
    \minprob{x}{p(x),}{q_i(x) \geq 0,}
\end{equation}
is known to be in $NP$ for generic polynomials $p,q_i$ \cite{Laurent2009}. This general form has as special cases many familiar (and difficult) optimization problems, such as binary linear programming and the max-cut problem. With the advent of efficient numerical algorithms for semidefinite programming at the turn of the century, approximate solutions to the polynomial optimization problem were suggested in the form of semidefinite relaxations. 

In particular, Laserre \cite{Lasserre_2001} provided a hierarchy of semidefinite programs which approximate to arbitrary accuracy the solution of \eqref{eq:polyopt}. This approach invokes results from real algebraic geometry and the theory of moments \cite{curto}. It regards the polynomial $p(x)$ as a linear combination of moments $y_\alpha = \int x^\alpha \mu(dx)$ of a positive measure $\mu$ defined on the set $\{q_i(x) \geq 0\}$. Consistency conditions on moments of positive measures imply semidefinite constraints on moment matrices:
\begin{equation}
    M \succeq 0; \quad M_{\alpha\beta} = \int x^{\alpha + \beta }\mu(dx)
\end{equation}
The correspondence between positivity of this (Hankel) matrix and the existence of the positive measure $\mu$ is the subject of the classical moment problems of Hamburger, Stieltjes, and Hausdorff \cite{Akhiezer_2020}. 

Laserre's hierarchy was subsequently generalized to problems in non-commuting variables, work which recognized potential applications to quantum mechanical systems \cite{Pironio_Navascues_Acin_2010}. The theory of non-commutative polynomial optimization was further developed in \cite{Araujo_Klep_Vertesi_Garner_Navascues_2023}, where the non-commutative analog of the Karush-Kuhn-Tucker optimality conditions were described. The methods were applied directly to spin systems to compute both the ground state energy and magnetization. While the commuting optimization problem \eqref{eq:polyopt} is $NP$-hard, the non-commuting problem can in general be ``Quantum-Merlin-Arthur" (QMA) hard. The complexity of this task is directly related to the known complexity of finding the ground state of an interacting spin chain, a class of problems which has been extensively addressed in the physics literature \cite{Kempe_Kitaev_Regev_2004,Ambainis_2013}. 

\subsubsection{Condensed matter and quantum information theory}
Contemporaneously with the mathematical developments discussed above, semidefinite methods were being applied in condensed matter theory to determine the ground state energies of various systems. 
Early work along these lines \cite{Nakata_Nakatsuji_Ehara_Fukuda_Nakata_Fujisawa_2001} studied fermionic systems and considered only two-body density matrices, which in the case of fermionic ground states of electronic systems, provides a decent variational ansatz. While this ``reduced density matrix" theory leveraged semidefinite programming, it was a variational approach rather than a bootstrap approach, with the variational parameters being matrix elements of the two-body reduced density matrix.

Some time later, Barthel \& Hubener and Mazziotti \cite{mazziotti,Barthel_Hubener_2012} systematized the semidefinite approach, defining a hierarchy of problems of the form \eqref{eq:btsp_energy_prob} and providing lower bounds on the energy of various spin systems. In \cite{Barthel_Hubener_2012}, considering any spin system algebra as being comprised of Jordan-Wigner fermion ladder operators, the authors consider a flag of subspaces given by (the span of) normal-ordered monomials of ladder operators of increasing degrees $k,k+1,\ldots$. Each of these subspaces defines, after taking expectation values, a set of e.g. $k$-point Green's functions, and the hierarchy of such subspaces provides successively better estimates of the ground state energy. 

While the ground state energy problem is already QMA-hard, Barthel \& Hubener note that consideration of the Green's functions is related to another QMA-hard problem, the so-called representability problem. Given some $k$-point operator $\OO^{(k)} = a_{i_1}^\dagger\cdots a_{i_m}^\dagger a_{i_{m+1}}\cdots a_{i_k}$, the exact $k$-point Green's function is given by $G_k(\OO) = \tr(\rho\OO^{(k)})$ where $\rho$ is the exact density matrix of the entire system. Given access only to ostenisble numerical values of $G_k(\OO)$ for various operators $\OO$, determining whether these values are consistent with being derived from a global physical state $\rho$ is QMA-complete. 

To evade this, the authors note that the condition $G_k(\OO) = \tr(\rho\OO^{(k)})$ for a global state is not equivalent to but requires that
\begin{equation}
    G_k(X^\dagger X) \geq 0
\end{equation}
for any operator $X$. Not all positive semidefinite $\OO_+$ operators have such a representation $\OO_+ = X^\dagger X$, but any such operator $X^\dagger X$ is positive semidefinite. The relationship between positivity and `sum-of-squares' representations is the link between the present methods and the content of Hilbert's 17th problem \cite{Laurent2009}, which famously asked whether positive polynomials admit sum-of-squares representations. 

Subsequent works \cite{Haim_Kueng_Refael_2020,Cho_Sun_2023} construct different hierarchies of operator bases but rely on the same positivity conditions and the notion of semidefinite relaxations; some introduce slightly different formulations of the ground energy problem which nonetheless provide lower bounds to the optimum \cite{Parekh_Thompson_2022,Westerheim_Chen_Holmes_Luo_Nuradha_Patel_Rethinasamy_Wang_Wilde_2023}. More recently, modifications of these types of semidefinite methods introduce aspects of renormalization \cite{Kull_Schuch_Dive_Navascues_2022} and clustering of operators \cite{Lin_Lindsey_2022}. In these works, the authors are able to compute the ground state of numerous spin systems to great accuracy including the Hubbard model, which, due to the fermion sign problem, is known to be difficult for Monte Carlo methods. 

The central focus of all these works is the computation of the ground state energy. While technically a QMA problem, many results (including ours) indicate that especially for translation-invariant systems, semidefinite methods quite quickly and easily give good estimates of the ground state energy, i.e. within a few percent error, even in relatively naive setups. However, the computation of ground state correlations via semidefinite methods is much less studied. In \cite{Haim_Kueng_Refael_2020}, the authors demonstrate that semidefinite methods give values of the spin-spin correlation function which qualitatively agree with density matrix renormalization group (DMRG) results. However, qualitative agreement is insufficient if one's goal is the accurate extraction of conformal data via this method. 

The topic of computing and bounding correlations by semidefinite or bootstrap methods has recently received some specific attention. The review \cite{Tavakoli_Pozas-Kerstjens_Brown_Araujo_2023} addressess a number of problems in quantum information and gives a good overview of how semidefinite and related methods can be applied to qubit systems. A pair of papers by Fawzi et al \cite{Fawzi_Fawzi_Scalet_2023a,Fawzi_Fawzi_Scalet_2023b} go a step further and provide `certified' algorithms for bounding expectation values of arbitrary observables in the ground state of a quantum system potentially at nonzero temperature. These works provide a concise and complete description of the relevant optimization problems and associated constraints, and even offer some proofs of the exponential convergence of these methods observed in certain cases. They address directly the issue of correlation functions and use their algorithm to produce rigorous bounds for the ground state correlators. However, for the size of optimization problem constructed, the obtained bounds are again far too weak for the analysis of conformal data, as we verified in our own numerical experimentation.

Recently, the classical Ising model in multiple dimensions was addressed by semidefinite methods in \cite{Cho_Gabai_Lin_Rodriguez_Sandor_Yin_2022}, where the authors leverage representation theory to block-diagonalize the semidefinite constraints and save computational overhead. They are able to place bounds on one-point and short distance two-point correlation functions; their investigations already demonstrate the difficulty of placing sharp bounds on longer-distance correlators at criticality. However, being an analysis of the classical model, the symmetries they utilize are not manifest in the quantum description we study at present. 

\section{The transverse-field Ising model}\label{sec:TIM}
To study the issue of determining correlations in spin systems by semidefinite methods, we now turn to the canonical example of a spin chain with a CFT limit: the transverse-field Ising model. It is a one-dimensional (plus time) spin chain on $N$ sites. The operator algebra $\A$ has $4^N$ elements: the four Paulis $\{I,X,Y,Z\} = \{I,\sigma^a\ |\ a=x,y,z\}$ at each site and their tensor products. They obey the standard algebraic relations
\begin{equation}
    [\sigma_i^a,\sigma_j^b] = 2i\varepsilon^{abc}\sigma_j^c \delta_{ij}\ \qquad(\text{no sum over }j).
\end{equation}
We use the Hamiltonian
\begin{equation}
    \label{eq:isingHam}
    \hat{H} = -\sum_{j=1}^N \left[X_j X_{j+1} + hZ_j\right]
\end{equation}
with periodic boundary conditions $N +1 \equiv 1$. The Hamiltonian is translation-invariant which means the ground state is translation invariant (in particular, $\mathbb{Z}_N$ invariant). We will use this symmetry to reduce the number of optimization variables. As is well known, in the thermodynamic limit $N \to \infty$ at $h =1$, the theory describes a free fermion compactified on the circle. This conformal field theory has central charge $c = 1/2$.

A specific program is specified by an operator basis $\B$, which we view as a vector of operators. We will consider a number of different operator bases and observe each of their convergence properties, but we begin with a very simple choice: define
\begin{equation}
    \B_0 = \{I\}\cup\{X_i,Y_i,Z_i\ |\ i = 1,\ldots,N\}.
\end{equation}
The moment matrix $M(\B_0)$ is constructed as
\begin{equation}
    M(\B_0) = \ex{\B_0 \B_0^\dagger} \succeq 0,
\end{equation}
and is therefore a $(3N + 1) \times (3N + 1)$ matrix for this particular choice of operator basis. Its matrix elements are correlation functions which will act as the optimization variables of the semidefinite program. Many of these matrix elements are linearly dependent either due to algebraic relations between the operators in $\B_0$ or by invariance of the functional $\ex{\cdot}:\A\to \RR$ under translation and potentially other discrete symmetries. 

To handle these equalities, we write the relevant semidefinite program in the linear matrix inequality (LMI) form, which is customarily regarded as the dual formulation. This regards the matrix $M(\B_0)$ as a linear combination of matrices, i.e.
\begin{equation}
    M(\B_0) = F_0 + x_1F_1 + x_2F_2 + \cdots
\end{equation}
where the $F_i$ are numerical, Hermitian matrices, and the $x_i$ are unique correlation functions: $\ex{X},\ex{X_1Y_3},\ex{Z_1Z_2}$, etc. For a given operator basis, the construction of the $x_i,F_i$ is independent of the objective function. In practice, this means that semidefinite solvers can cache matrix factorizations, leading to increased speed at runtime. 

\subsubsection{Computational implementation}
To flexibly construct many different semidefinite programs, we used the \verb|OpenFermion| package from Google Quantum AI \cite{McClean:2017ims}. This Python package defines efficient Pauli and fermionic algebra objects and includes routines for exact diagonalization, Trotterization, etc. 

Operators in $\A$ form objects which are linear combinations of products of Pauli matrices. Products are then encoded as $N$-length strings: $\ex{X_1} = XIII\cdots I$, $\ex{Z_1Z_3} = ZIZII\cdots I$, and so on. To take advantage of translation invariance, we define a cyclic invariant hash function which takes as input a string of length $N$ and outputs an integer hash value invariant under cyclic shifts of the string. In general, the hash function should be constructed to be invariant under symmetries of the ground state. Each unique hash value corresponds to the orbit of a given correlation function under $\mathbb{Z}_N$ translation acting on the operators and becomes an optimization variable $x_i$ of the semidefinite program.  The number of such (dual) variables $n_d$ depends on the number of sites, the length of the operator basis, and any special algebraic relations that may exist between elements of the basis. Operator bases must be chosen so that $\ex{\hat{H}}$ may be written as a linear combination of the optimization variables $x_i$. 

A semidefinite program for any spin chain ground state is therefore specified by a Hamiltonian $\hat{H}$ and an operator basis $\B$; the algebraic relations are automatically computed to determine the $x_i$. Finally, one defines the optimization problem 
\begin{equation}
    \label{eq:bootstrapSDP}
    \minprob{x_i}{\ex{\hat{H}} = \sum_{i =1}^{n_d} c_ix_i}{M(\B_0) = F_0 + \sum_{i=1}^{n_d} x_iF_i \succeq 0 }
\end{equation}
At this point, one may choose to add extra constraints. Outside of enforcing symmetries like cyclic or translation invariance on the expectation values, there are two types of constraints one can generically add to the ground-state problem. Both types are considered and utilized in the ``certified" algorithms of Fawzi et al \cite{Fawzi_Fawzi_Scalet_2023a,Fawzi_Fawzi_Scalet_2023b}. The first are eigenstate constraints of the form
\begin{equation}
    \ex{[\hat{H},\OO_i]} = 0.
\end{equation}
These apply in any energy eigenstate, and essentially state that time evolution for the states in question is trivial. The second are unique to the ground state and state that any perturbation to the state raises the energy:
\begin{equation}
    \ex{\OO_i^\dagger [\hat{H},\OO_i]} \geq 0
\end{equation}
We experimented in multiple instances with including these constraints in the optimization, but find that they rarely affect the outcome for SDPs written in terms of spin variables. This behavior was also noted by Lawrence \cite{Lawrence:2021msm}. However, the equality constraints appear to be crucial for constructing fermionic SDPs. Note that the inclusion of these constraints may require new operators to be added to the basis (and therefore more optimization variables), which can in practice diminish the performance of the solver. 

Once the problem is constructed in this fashion, we use the the \verb|cvxpy| package with MOSEK to solve the SDP \cite{agrawal2018rewriting,mosek}. While we experimented with using sparse matrix data structures, we did not find that these increased the performance and sometimes even slowed down the internal compilation of the SDP. This is likely a function of how \verb|cvxpy| handles the problem compilation, and sparse matrices are likely to help if interfacing directly with an SDP solver. 

\subsubsection{Choosing an operator basis}
For a fixed Hamiltonian and system size $N$, the computational implementation of the bootstrap takes in an operator basis $\B$ and outputs numerical approximations of the energy and any ground-state correlation functions contained in the (optimized) moment matrix $M(\B)$. These can be numerically extracted and subsequently compared to the known, exact values. 

To understand the performance of the method, we consider four different operator bases. The first is the basis of one-site operators $\B_0$ with size $3N + 1$. To go one step up in the hierarchy, we define
\begin{equation}
    \B_1 = \B_0 \cup \{ \sigma^a_1\sigma^b_2\ |\ a,b = x,y,z\}
\end{equation}
So far, the hierarchy $\B_0 \subset \B_1$ is similar to that proposed in \cite{Haim_Kueng_Refael_2020}, but at level $\B_1$ we anchor each nearest-neighbor two-point function at site 1, resulting in a basis of size $|\B_1|= 3N + 10$. At the proximate level, consider all nearest-neighbor two-point functions, taking
\begin{equation}
    \B_2 = \B_0 \cup \{ \sigma^a_1\sigma^b_n\ |\ a,b = x,y,z;\ n = 2,\ldots,N \}
\end{equation}
This basis is much larger, with size $|\B_2| = 12N + 1$. This corresponds to the second level in the hierarchy of \cite{Haim_Kueng_Refael_2020}. 

Finally, we consider a special basis: that of the Jordan-Wigner fermion ladder operators. Recall that the JW fermion operators $c_i, c_i^\dagger$ are constructed of long strings of Paulis: 
\begin{align}
    c_i &= \frac{1}{2}\prod_{j=0}^{i-1}\left[-Z_j\right] (X_i - iY_i)\\
    c_i^\dagger &= \frac{1}{2}\prod_{j=0}^{i-1}\left[-Z_j\right] (X_i + iY_i)
\end{align}
We use these operators to define the fermion basis
\begin{equation}
    \B_F = \B_0 \cup \{c_i, c_i^\dagger\ |\ i = 1,\ldots,N\}
\end{equation}
where the basis $\B_0$ is included so that all two-point correlation functions of interest can be obtained from the program optimum.  Since the Ising Hamiltonian is known to be quadratic in this basis, we should expect that the fermion basis yields the best results, similarly to the harmonic oscillator system in the Schrodinger bootstrap \cite{Berenstein:2021loy}. Of course, its introduction here relies on our prior analytical knowledge that the system has an exact free-fermion description. 

While these operators obey fermion algebraic relations, they are still fundamentally written in terms of spin variables. One can instead construct SDPs which use the JW fermions as algebraic primitives. While this approach can work well to determine the ground state, it becomes very difficult to extract local spin correlation functions from the fermion description. For interacting fermionic Hamiltonians, the two-point spin correlation functions are necessarily highly non-local in the JW fermions, making fermionic primitives impractical for estimating correlation functions from optimality.

\subsubsection{Ground state energy}
We begin by considering the Ising Hamiltonian \eqref{eq:isingHam} for a range of values of the transverse field $h$. The bases $\B_0,\B_1,\B_2$ contain operators which capture more local dynamics of the spin chain; the fermion basis $\B_F$ instead contains operators which are very non-local in the sense that they involve Pauli operators over many sites of the system. 
\begin{figure}[!h]
    \centering
    \includegraphics[width = 0.8\columnwidth]{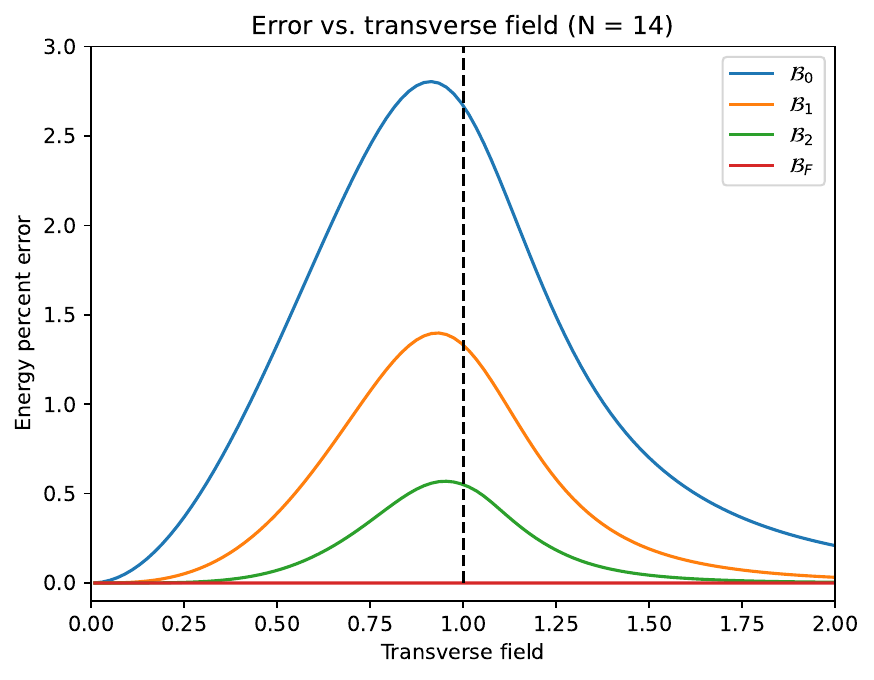}
    \caption{Percent error vs. transverse field for the four operator bases described in the previous subsection. The critical point $h = 1$ is demarcated. }
    \label{fig:Ising_energy_error}
\end{figure}
At criticality $h = 1$, the system contains strong long-range correlations. It is therefore unsurprising that the bases constructed of local operators have their largest percent errors near the critical point. This is reflected in Fig. \ref{fig:Ising_energy_error}, shown for a system with $N = 14$ sites, where the fermion basis results in an essentially exact answer, likely deviating by only floating-point and solver error. 

Note that even the simplest basis $\B_0$ already gives an approximate ground state energy within a few percent error of the known, exact value. Since $\B_0 \subset \B_1 \subset \B_2$, one is guaranteed that the approximated energy increases in fidelity as one moves up the hierarchy. However, these increases may be of vastly differing magnitude relative to the increase in the size of the basis and the associated computational time. While the bases $\B_0,\B_1$ define SDPs solved in only a matter of seconds, the bases $\B_2,\B_F$ may take multiple minutes to reach optimality. The time taken to solution versus the system size $N$ is tracked below in Fig. \ref{fig:Ising_time_nsites}.
\begin{figure}[!h]
    \centering
    \includegraphics[width = 
    0.8\columnwidth]{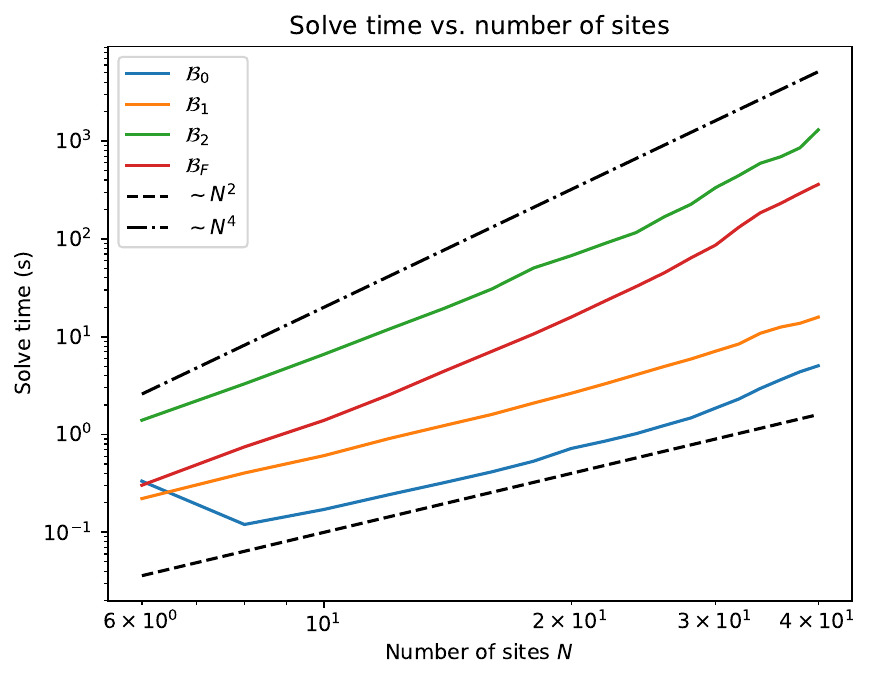}
    \caption{Time to optimality for each operator basis and over a range of system sizes; scaling comparison curves are included for solve time proportional to $N^2,N^4$. }
    \label{fig:Ising_time_nsites}
\end{figure}
Regardless of the basis under consideration, the size of the basis grows linearly in the system size $N$, and the number of dual variables grows at worst like $N^3$. At a fixed value of the transverse field, we observe that the energy error, as function of the system size, plateaus quickly (around $N \sim 20$) for each basis to a constant value. 

These results confirm what has been found numerous times in the literature: the bootstrap method, a set of semidefinite relaxations, yields good accuracy in predicting the finite-size ground state energy of the spin chain even for small operator bases at most linear in the size of the system $N$. We note that many previous works make very detailed choices of operator bases; experimentally, we find generically that one and two-point operators do quite well to estimate the energy of translation-invariant nearest-neighbor Hamiltonians.

\subsubsection{Magnetization}
An important question for numerical methods in many-body systems is whether they can reliably detect phase transitions. At the quantum critical point $h =1$, the transverse-field Ising model displays a second-order phase transition; the magnetization $\ex{Z}$ has a discontinuous second derivative $\partial_h^2\ex{Z}$ with respect to the transverse field $h$. In any finite-size system, there is technically no phase transition: the second derivative of the magnetization should display a large but finite spike at the quantum critical point. 
\begin{figure}[!h]
    \centering
    \includegraphics[width = 0.8\columnwidth]{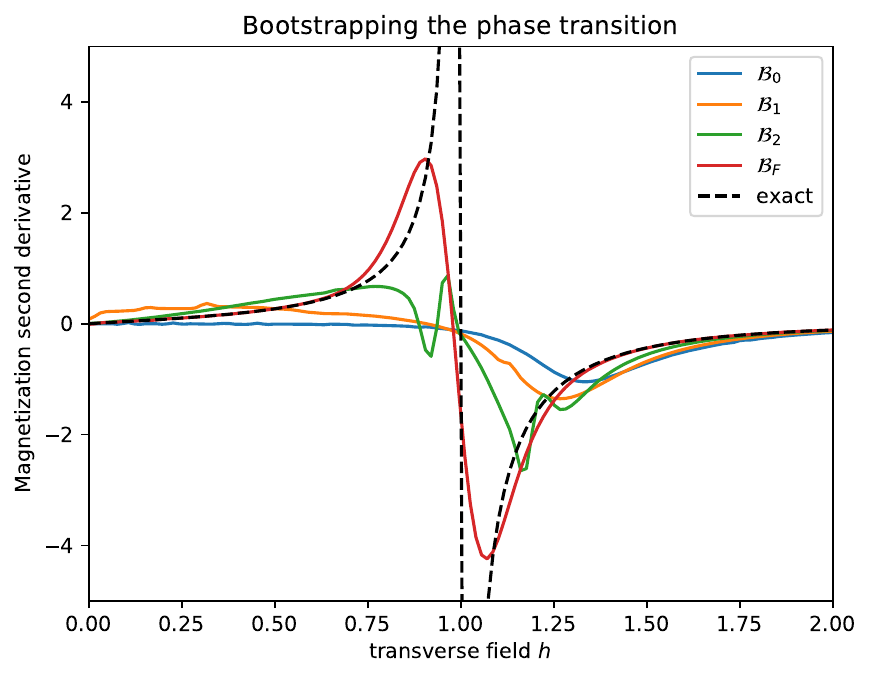}
    \caption{The second derivative of the magnetization $\partial^2 \ex{Z}/\partial h^2$ with respect to the transverse field $h$, with the exact curve shown in black, dashed. The fermion basis detects the divergence, but the other bases show much less sensitivity. }
    \label{fig:sdp_phase_mag}
\end{figure}

Computing the magnetization by SDP methods is the first test of the bootstrap's ability to compute correlation functions. Since the magnetization enters directly into the Hamiltonian, whereas most two-point functions serve effectively as slack variables, the expectation should be that the bootstrap performs well in computing this quantity. 

In Fig. \ref{fig:sdp_phase_mag}, we compute the second derivative of the magnetization $\partial^2\ex{Z}/\partial h^2$ exactly and by SDP with the four operator bases introduced earlier. The fermion basis clearly detects the phase transition, and the complete two-point function also displays some sensitivity. Most bases display a distinct asymmetry about the critical point, a behavior also evident in the energy error versus transverse field of Fig. \ref{fig:Ising_energy_error}. 

\subsubsection{Correlation functions}
\begin{figure*}[!t]
    \centering
    \includegraphics[width = \columnwidth]{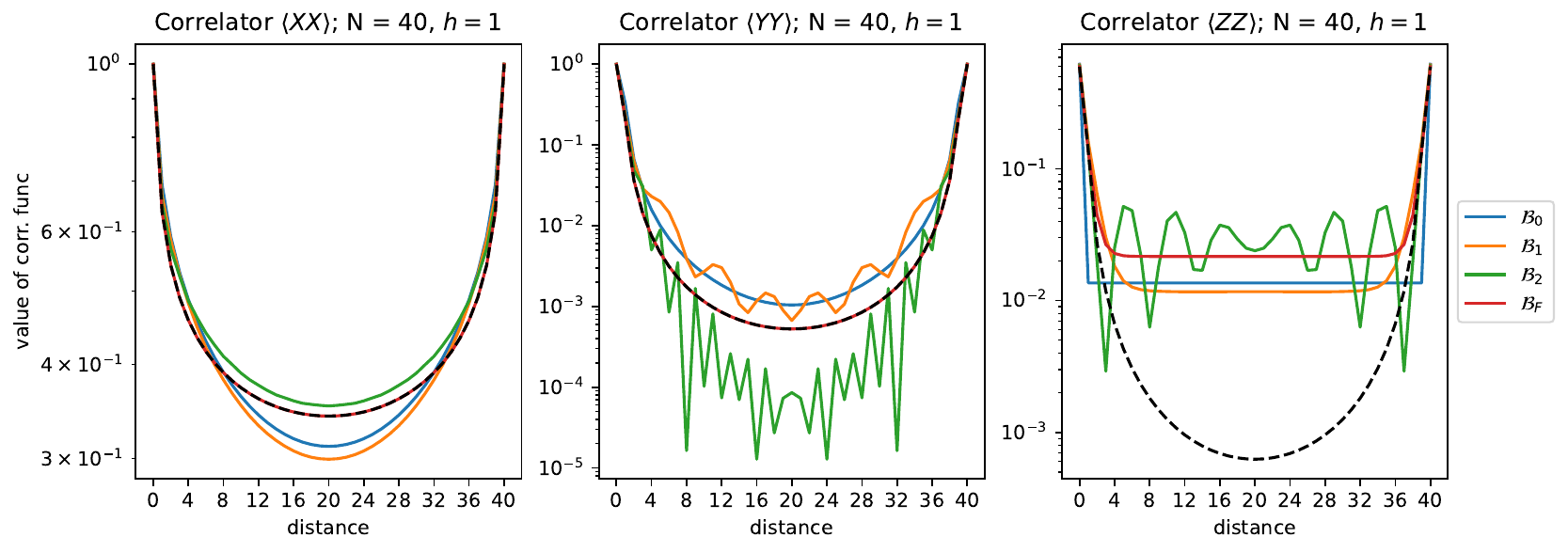}
    \caption{Correlation functions (in particular, their absolute value) as estimated from the optimal values of the SDP, compared with the exact expressions (black, dashed). While basic qualitative agreement is achieved for the $XX$ correlator, the estimates of the others are not good. Note that the small fluctuations are magnified by the logarithmic scale.}
    \label{fig:corrcompare}
\end{figure*}
We turn now to the spin-spin correlation functions. With the goal of determining conformal scaling dimensions from the critical spin chain, we must study the connected correlation functions $\ex{X_1X_{1+n}}^c,\ex{Y_1Y_{1+n}}^c,\ex{Z_1Z_{1+n}}^c$ at criticality as functions of $n$. For the transverse-field Ising model, exact expressions for these quantities at finite $N$ are known and are reproduced in the Appendix. For each of the bases $\B_0,\B_1,\B_2,\B_F$ under consideration, we compute the exact correlation functions and compare these to the approximate values obtained at optimality of the program \eqref{eq:bootstrapSDP}. These comparisons are pictured in Fig. \ref{fig:corrcompare}.

The results make it clear that, as previously observed in \cite{Haim_Kueng_Refael_2020,Fawzi_Fawzi_Scalet_2023a}, the SDP relaxations can give values for the correlation functions which \textit{qualitatively} agree with the exact results. Small deviations in the values are magnified by the logarithmic scale on the plot. However, the qualitative agreement of Fig. \ref{fig:corrcompare} is in general not sufficient to obtain accurate results for the scaling dimensions, which depend in a detailed way on the large separation region. In this regime, the expectation values are small, but not small enough. 

\begin{figure*}[t]
  \centering
  \begin{minipage}{\linewidth}
    \centering
    \includegraphics[width=\textwidth]{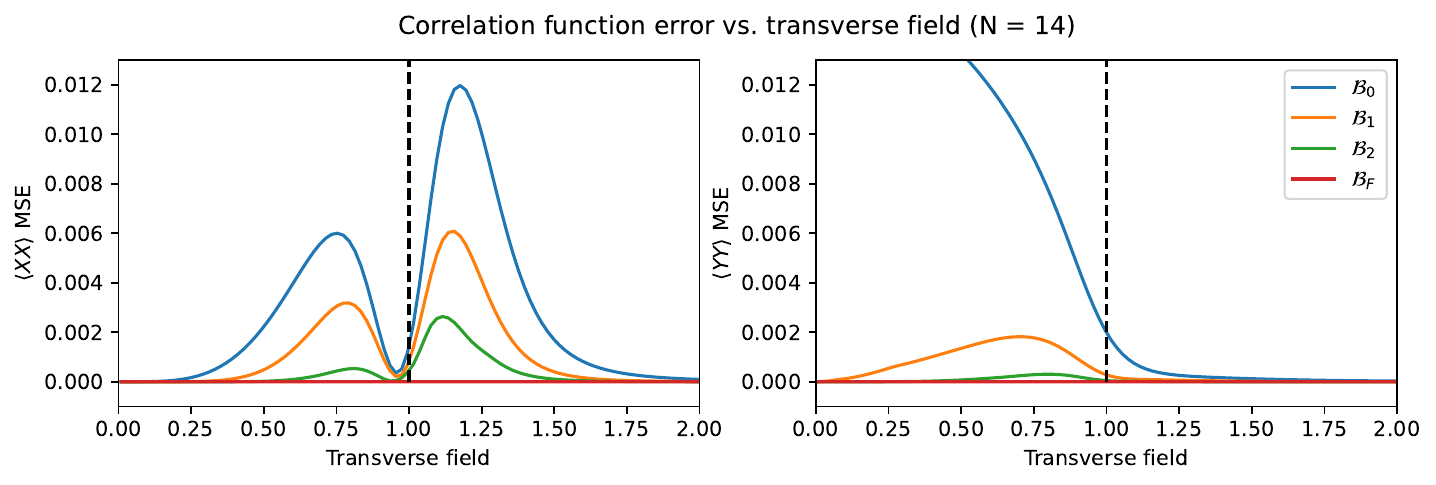}
    \caption{Correlation function mean squared error (MSE) versus transverse field for the $\ex{XX},\ \ex{YY}$ correlation functions. While each basis in the hierarchy improves the overall error, the error profile across phases of the model is much more nontrivial than the error profile observed for the energy estimate.}
    \label{fig:corr_error_TF}
  \end{minipage}

\end{figure*}
To study in more detail the error associated to the correlation function estimates, we can consider the mean-squared-error of the connected correlators $\ex{XX},\ex{YY}$ for each basis versus the transverse field $h$, shown in Fig. \ref{fig:corr_error_TF}. We compute the squared error e.g.  $|\ex{X_1X_{1+n}}_{\text{exact}} - \ex{X_1X_{1+n}}_{SDP}|^2$ and average over all distances $n$. The nontrivial error profile shows that convergence to the correct correlation functions happens in less predictable fashion compared to the error seen in the energy estimates and, interestingly, the $\ex{XX}$ correlator has a local minimum in the error near the quantum critical point. The correlation error also approaches zero in every basis as we move into the ordered phase $h \to \infty$.   


\subsubsection{Determination of conformal data}

To extract conformal dimensions from the $N \to \infty$ limit of the critical spin chain, we have to compare our numerical estimates to correlation functions of the 2D CFT compactified on the cylinder. For a conformal field $\phi(z,\bar{z})$ on the complex $z$ plane, the scaling dimension $\Delta_\phi=h+\bar{h}$ is defined by the two-point function as
\begin{equation}
    \left\langle\phi(z, \bar{z}) \phi\left(z^{\prime}, \bar{z}^{\prime}\right)\right\rangle\sim(z - z')^{-2h}(\bar{z}-\bar{z}^\prime)^{-2\bar{h}}
\end{equation}
Following Cardy \cite{Cardy_1986}, we consider the conformal mapping $w = (L/2\pi) \log z$ which takes the plane to a finite strip of width $L$. We are primarily interested in the equal-time correlators; with $w = t + i\sigma$ with $\sigma \in [0,L]$, the correlation function at two equal-time points $(\sigma_1,\sigma_2)$ is given by
\begin{equation}\label{eq:correlatorstripL}
    \ex{\phi(\sigma_1)\phi(\sigma_2)} \sim \sin\left(\frac{\pi (\sigma_2 -\sigma_1)}{L}\right)^{-2\Delta_\phi}
\end{equation}
Therefore, to use the (finite) spin-chain correlation functions to compute $\Delta_\phi$, one should regress to the expression
\begin{equation}\label{eq:fitdelta}
    \ex{X_1X_{1+n}} \sim \sin(\pi n/N)^{-2\Delta_X}
\end{equation}
Strictly speaking, the critical spin chain/CFT correspondence is only valid in the $N \to \infty$ limit. Therefore, one should investigate the finite-size correlation functions at large system size $N$ and with $n \gg 1$, where the long-distance behavior is more reliably reflective of the physics of the conformal theory. In practice, this means using a short-distance cutoff and only considering values of the correlation function at $n \gtrsim 5$ for system sizes of order $N \sim 40$. 
\begin{table}[]
    \centering
    \begin{tabular}{c||c|c}
         basis & exponent estimate & \% error\\
         \hline
         $\B_0$ & 0.2008 & 60.661\\
         $\B_1$ & 0.2133 & 70.712\\
         $\B_2$ & 0.1404 & 12.3439\\
         $\B_F$ & 0.1248 & 0.1994
    \end{tabular}
    \caption{Estimates of the critical exponent $\Delta_X = 1/8 = 0.125$ from the SDP correlation functions for each basis and the associated percent error, computed here with $N = 40$ and a cutoff of 5 sites on each side.}
    \label{tab:isingExps}
\end{table}

The results of this procedure for the $\ex{X_1X_{1+n}}$ correlator are included in Table \ref{tab:isingExps} with $N = 40$ and considering $5 \leq n \leq 35$. As is well known, with the Hamiltonian \eqref{eq:isingHam} at criticality the $X$ operator becomes the spin field with conformal dimension $\Delta_X = 0.125 = 1/8$ in the thermodynamic limit. From the results, it is clear that the fermionic basis provides the best estimate of the scaling dimension. When these estimates are extracted only by regressing against the SDP correlation functions, there is no way to extract rigorous error bounds or even confidence intervals, at least from a single run of the program. Interestingly, the error in the estimates from bases $\B_0,\B_1$ increases with the system size $N$, even when modulating the cutoff. 

The correlation functions and the energy also permit an estimation of the central charge of the conformal theory. The ground-state energy of a critical spin chain has well-known finite-size scaling properties with respect to the system size $N$ \cite{Cardy_1986,henkel1987finite}. In particular, one has
\begin{equation}
    e_0^{(N)} \equiv E_0^{(N)}/N \sim e_0 - \frac{\pi c}{6N^2} + \mathcal{O}(N^{-3})
    \label{eq:cardy_scaling}
\end{equation}
where $e_0$ is the ground state energy density per site in the thermodynamic limit $N\to\infty$ and $c$ is the central charge. In general, the dimensions in Eqn. \eqref{eq:cardy_scaling} are not exactly correct. That is, they depend on the speed of light in the lattice theory; we call the lattice speed of light $v$ by abuse of notation. In general, one must do a normalization step to compute the effective speed of light in order for the parameter $c$ in \eqref{eq:cardy_scaling} to be equal (in dimensions and value) to the CFT central charge. An example of this type of normalization procedure is handled in \cite{Milsted_Vidal_2017}. 

Given an estimate of the critical exponent $\Delta_X$, we can directly estimate the effective speed of light from the numerical SDP. The procedure goes as follows. For an operator with scaling dimension $\Delta$, the unequal time thermodynamic limit correlation function depends on the speed of light $v$ as
\begin{equation}
    \left\langle\mathcal{O}(0, t) \mathcal{O}(x, 0)^{\dagger}\right\rangle\propto\frac{1}{\left(x^2-v^2 t^2\right)^{\Delta}}
\end{equation}
The idea is to take time derivatives of this expression to extract factors of $v$ and regress an estimate for the speed of light directly from the correlators; the details of the manipulations are worked out in the appendix. For an operator on a lattice of size $N$, one finds
\begin{equation}
    \label{eq:centralchargeratio_ising}
    \frac{\langle\ddot{\mathcal{O}}(0,0) \mathcal{O}(x, 0)^{\dagger}\rangle}{\left\langle\mathcal{O}(0,0) \mathcal{O}(x, 0)^{\dagger}\right\rangle}=\frac{-2 \pi^2 v^2 \Delta}{N^2 \sin (\pi x / N)^2},
\end{equation}
where we consider a ratio of correlation functions so that any unphysical normalization of the field/operator does not enter into this expression. To evaluate the time derivative on the left hand side we can use the algebraic structure of the time evolution: $\ddot{\mathcal{O}} = -[\hat{H},[\hat{H},\mathcal{O}]]$. 

Let us carry out this procedure for the $X$ field with the Hamiltonian \eqref{eq:isingHam} at criticality. From the operator algebra one finds
\begin{equation}
    \label{eq:threept_speedoflight}
    \ex{\ddot{X}_1X_{1+n}} = 4\ex{X_1X_{1+n}} - 4\ex{Z_1X_2X_{1+n}} - 4\ex{Z_1X_NX_{1+n}}
\end{equation}
We must choose an operator basis which will give us access to numerical estimates of these three-point functions. To this end, we use an extension of the fermion basis $\B_X = \B_F \cup \B_2$ and solve the resulting SDP on a chain of length $N=20$; this problem already has $\sim 7.5$k optimization variables and takes about 60 seconds to solve. From the result, we first regress the connected $\ex{XX}$ correlator to obtain an estimate of the scaling dimension $\Delta_X$; we find $\Delta_X \simeq 0.12412$.

Next, we compute the SDP estimates of the ratio $-\ex{\ddot{X}_1X_{1+n}}/\ex{X_1X_{1+n}}$ over distances $3 \leq n \leq 17$, using a short-distance cutoff to ensure we are in the regime of validity of the expression \eqref{eq:centralchargeratio_ising}, and regress to the right-hand side of \eqref{eq:centralchargeratio_ising} to determine $v$. With our scaling dimension estimate, we find $v \simeq 2.02$. This is in line with the known analytical results, where the fermionic representation of the Ising Hamiltonian gives a dispersion relation with exact lattice speed of light $v_{exact}=2$. In this case, one technically need not regress the entire expression against $x$, but doing so ensures that the results should replicate for any given value of the position. 

Finally, we can use all these estimates to address the issue of the central charge. The dimensionally consistent finite-size scaling law says that for a lattice of size $N$, one has at leading order
\begin{equation}
    e_0^{(N)} \sim e_0-\frac{\pi cv}{6 N^2}. 
\end{equation}
Using the same basis $\B_X$, we solve the SDP over a range of system sizes $N = 12,\ldots 20$ and regress the energy density estimates to the scaling law above, using our estimated speed of light. Doing so yields $e_0 \simeq -1.27323$ and $c \simeq 0.4967$. Comparing to the exact values, the estimate of the thermodynamic energy density is exact to less than 1 part in $10^{-5}$ and the central charge estimate is within 1\% error. 

The analysis of this section demonstrates that in principle, SDP relaxations of exact ground state problems can be used to estimate scaling dimensions and central charges of CFTs arising from critical spin chains. However, doing so with high fidelity depends strongly on one's choice of operator basis. In numerical experiments, we found that using composite bases like $\B_0 \cup \B_F$, $\B_1 \cup \B_F$ instead of the large basis $\B_X = \B_2 \cup \B_F$ gave essentially unusable results for the speed of light regression (the difficult part of the procedure), despite furnishing good estimates of the scaling dimension $\Delta_X$. For non-integrable models, determining a sufficiently good operator basis may be extremely difficult, and may push the memory requirements of the semidefinite solver beyond what is reasonable for laptop-based computation.  

\subsection{Non-invertible symmetries}
At criticality, the transverse-field Ising model exhibits a non-invertible symmetry: the Kramers-Wannier duality. This strong/weak duality transformation maps the transverse field term to the nearest-neighbor term in the Hamiltonian. In particular, it implies the following equality of expectation values at criticality:
\begin{equation}
    \ex{X_jX_{j+1}} = \ex{Z_j}.
\end{equation}
This duality is part of the physics but is in no way specified or enforced in the construction of the SDP. By studying how these correlation functions approach or diverge from one another near the critical point, we can better understand how and if the SDP can detect non-trivial aspects of the physics. 
\begin{figure}
    \centering
    \includegraphics[width = 0.8\columnwidth]{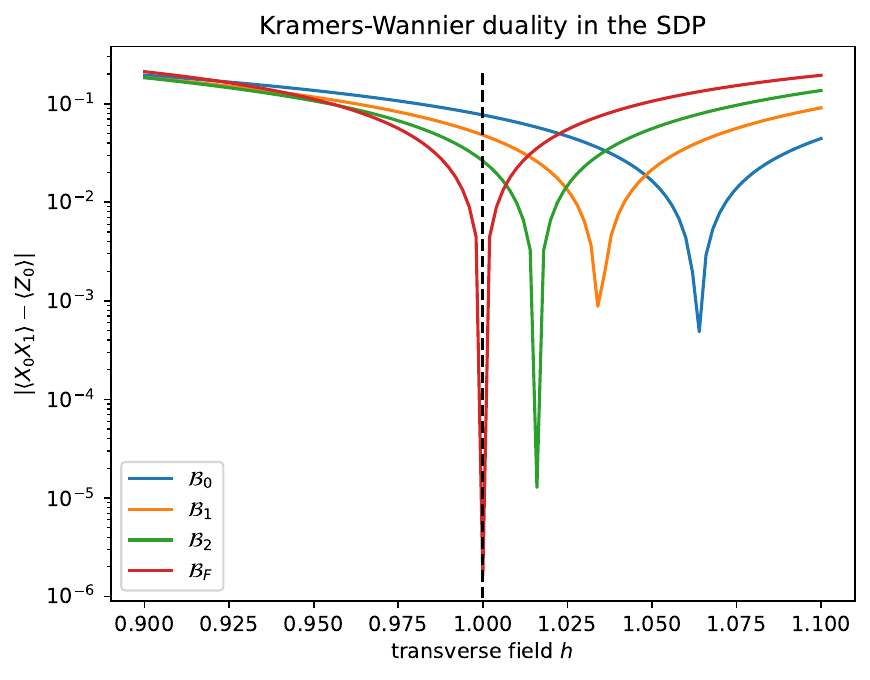}
    \caption{Testing whether Kramers-Wannier duality holds in the Ising model SDP at optimality. Only the fermion basis obtains the correct result. Interestingly, the other bases seem to approximately reach the dual point at values of the transverse field slightly away from criticality. As the bases grow in size, the self-dual point approaches the quantum critical point, denoted in black.}
    \label{fig:kw_duality}
\end{figure}
To test whether the SDP results respect this symmetry, we plot in Fig. \ref{fig:kw_duality} the absolute difference between the values of these correlators as obtained at optimality for each operator basis considered here and across a small range of values of the transverse field centered around the critical point $h = 1$.

The duality works as follows:  first, fix the subalgebra of operators $K_{2j}=X_{j}X_{j+1}$ and $K_{2j-1}=Z_j$ (that is, do not include the individual $X_j$).
With this construction, the individual $K$ anticommute with their nearest neighbors and commute with the rest of the elements. The subalgebra generated by the $K_n$ has an automorphism $K_n\to K_{n+1}$ that leaves the Hamiltonian invariant at criticality; this is the non-invertible symmetry. A two point function operator $X_j X_k$ is a non-trivial operator in the algebra of the $K$ defined by
\begin{equation}
X_j X_k = K_{2j} K_{2j+2} \dots K_{2k-2},
\end{equation}
which in the algebra of the $K$ is a non-local chain. When we use the duality, the bilocal operator gets mapped to the non-local chain
$K_{2j+1}\dots K_{2k-1}= Z_{j+1} \dots Z_{k}$.

Notice that if one uses only local and bilocal operators in the basis of correlators for the SDP, the operator basis for the optimization does not have the images of the operators under the duality. Expectation values of operators related to each other in this way by the Kramers-Wannier duality should be identical to each other and if the equations are sufficiently symmetric between the operators that are related to each other, the SDP solvers tend to find such symmetric solutions.
The equations for the solver can not see the symmetry directly: it is blind to the knowledge of the non-local operators in the local and bilocal basis. To ameliorate this problem, one needs to include non-local operators from the beginning, like $Z_{j+1} \dots Z_{k}$ perhaps with additional decorations at the endpoints. Indeed, the fermions obtained by the Jordan-Wigner transformation are optimal as they transform very simply into each other under the Kramers-Wannier duality and also allow one to write the Hamiltonian easily. This explains why only the fermion basis is able to see the duality in Figure \ref{fig:kw_duality}.

\subsection{Open boundary conditions in the bootstrap}
One nice feature of the bootstrap approach, noted and applied in multiple works \cite{Han:2020bkb,Lawrence:2021msm,Kazakov:2022xuh}, is that it may still theoretically be applied to systems in the thermodynamic limit $N = \infty$. In particular this is true for the spin systems we consider here. 

There are three classes of boundary conditions we can consider for the spin chain bootstrap and each have different symmetries. In practice, this means different sets of boundary conditions will cause different sets of expectation values to be identified when constructing the set of optimization variables. The first class of boundary conditions are periodic on a chain of length $N$ (which is what we have used up until this point) with Ising Hamiltonian
\begin{equation}
    \hat{H}_C = -\sum_{j=1}^N X_jX_{j + 1} - h\sum_{j = 1}^N Z_j;\quad X_{N+1} \equiv X_1. 
\end{equation}
With periodic boundary conditions, expectation values are grouped by their $\mathbb{Z}_N$ orbits, a $\mathcal{O}(N)$ reduction in the number of unique optimization variables. We can also consider the class of open boundary conditions on a chain of length $N$; the Hamiltonian becomes
\begin{equation}
    \hat{H}_O = -\sum_{j=1}^{N-1}X_jX_{j+1} - h\sum_{j=1}^N Z_j.
\end{equation}
Note that this system is not translation invariant, and so there is no immediate grouping of expectation values. Finally, we can consider the thermodynamic limit of the system $N \to \infty$. Here we have translation invariance and we use as the objective function the energy density, as in \cite{Lawrence:2021msm}:
\begin{equation}
    \hat{h}_j = -X_jX_{j+1} - hZ_j.
\end{equation}
Each of these boundary conditions define SDPs with vastly differing numbers of optimization variables. In particular, let $N = 20$ and consider the basis $\B_2$ consisting of all one-point operators and all nearest-neighbor two-point operators in a system of size $N$ (240 operators). The open BC SDP has 24720 optimization variables, the thermodynamic limit SDP has 4305 variables, and the periodic BC SDP has only 1242 variables. This results in an extreme difference in speed and memory requirements at runtime. 
\begin{figure}[!h]
    \centering
    \includegraphics[width = 0.8\columnwidth]{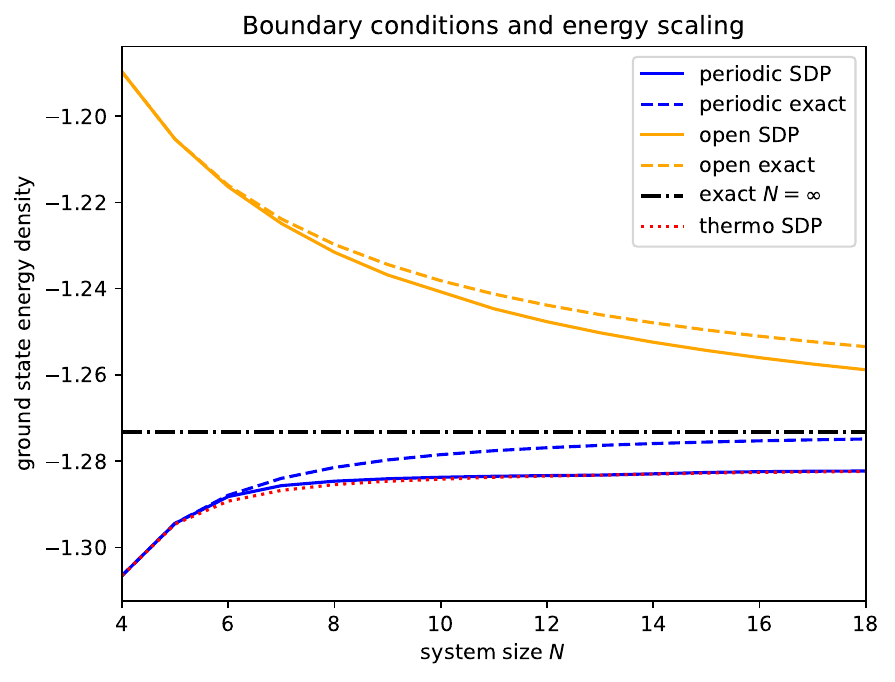}
    \caption{Energy density scaling with system size for Ising SDP results with periodic, open, and thermodynamic boundary conditions. Results from exact diagonalization are shown as dashed lines. Notably, the thermodynamic limit SDP appears to essentially reproduce the periodic SDP results.}
    \label{fig:bc_energy_scaling}
\end{figure}
Fig. \ref{fig:bc_energy_scaling} shows the results of solving these three classes of Ising SDP over a range of system sizes with the basis $\B_2$ (a function of system size). The results of exact diagonalization are shown as dashed lines. The open BC SDP finds the correct physical behavior, providing a good bound for the exact solution of the open system. However, the thermodynamic SDP appears to reconstruct the results of the periodic SDP and with a steeper computational cost. 
\begin{figure}
    \centering
    \includegraphics[width = \textwidth]{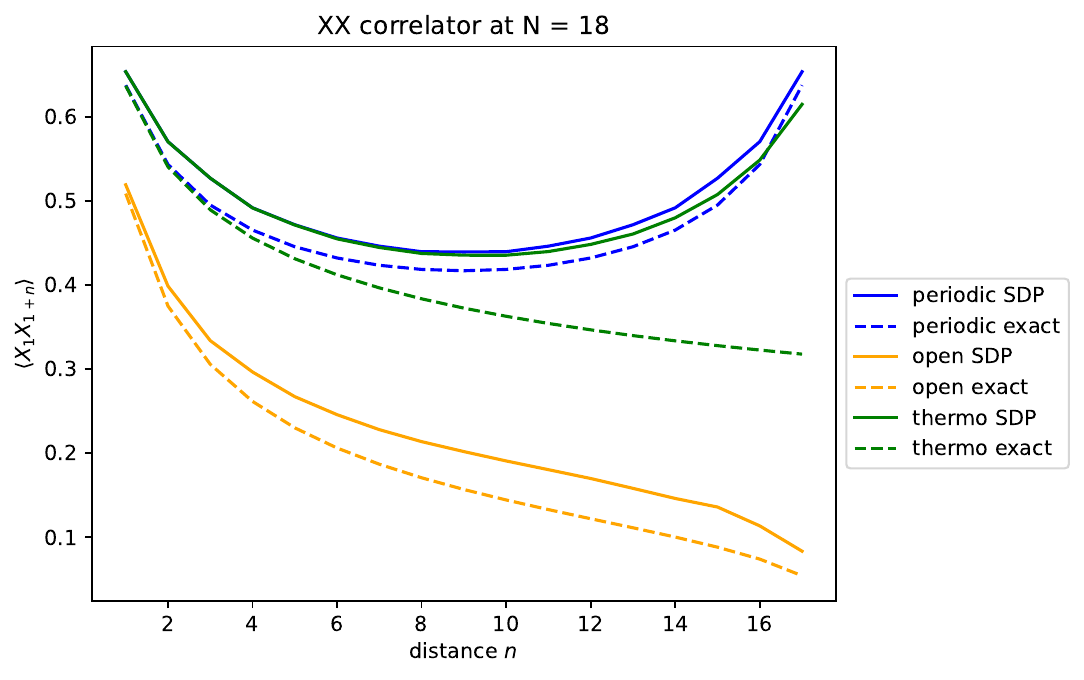}
    \caption{Comparison of $\ex{XX}$ correlation function values obtained for the basis $\B_2$ constructed for a system of $N = 18$ sites by each class of SDP: periodic, open, and thermodynamic. While the periodic and open results qualitatively reproduce the desired physical behavior, the thermodynamic SDP results appear to approximately (though not exactly) reconstruct cyclic invariance of the correlation function. }
    \label{fig:bc_xx}
\end{figure}

To understand better the convergence of the periodic and translation-invariant thermodynamic SDPs, we can compare the optimal values of the connected $\ex{XX}$ correlation function for the same basis at system size $N = 18$, shown in Fig. \ref{fig:bc_xx}. Notably, the translation-invariant thermodynamic SDP appears to reproduce a quasi-periodic behavior. To understand why this is, one must consider exactly what information the SDP sees about the system as defined by its objective function and constraints. 

In the periodic and thermodynamic SDPs, the objective functions are identical (up to a multiplicative factor) by translation/cyclic invariance. While the periodic SDP enforces cyclic invariance directly, the thermodynamic SDP sees only translation invariance. The only information the thermodynamic SDP has about the system `size' is contained in the basis of operators used to construct the moment matrix. 

The translation-invariant SDP enforces approximate translation invariance due to implicit symmetries in the way the problem is specified. Clearly, the physical results of any SDP should be invariant under certain permutations of the operator basis, which have no physical meaning. Permutations acting on the operator basis $\B$ induce an action by conjugation of the moment matrix $M = \ex{\B \B^\dagger}$. The moment matrix, constructed of expectation values, is only invariant under this conjugation if the permutation is compatible with the operator algebra and symmetries of the state functional (and therefore the algebraic relations between matrix elements of $M$). 

Consider a basis of $X_i$ for $i = 1,\ldots,N$ and $Y_i$ for sites $i=1,\ldots,K$. If $K = N$, we have an exact reordering symmetry $i \to N-i+1$. This symmetry (essentially space parity) leaves all translation-invariant moment matrix elements invariant, and so any optimal result of the SDP will also (approximately) have this symmetry. Therefore, even a problem without explicitly enforced cyclic symmetry may approach a cyclic symmetry at optimality as a consequence of discrete symmetries in the problem grammar. Testing this behavior with the basis just described, one sees that the translation-invariant SDP approximately finds periodicity for $K \geq N/2$, but does not find periodicity for $K < N/2$. That periodicity appears at $K = N/2$ is no accident and is a consequence of translation invariance coupled with the reordering symmetry. Future work, beyond the scope of this paper, should investigate the relationship between permutation actions on the operator basis and the symmetries obeyed (or enforced) on the linear functional used to build the moment matrix.

These results contrast existing statements that the thermodynamic SDP, with an $N$-finite basis, is computing an approximation of the system with open boundary conditions. The open system has no translation invariance due to the boundary conditions and hence receives a completely different objective function, leading to the physically expected result. In contrast, exactly what system the thermodynamic SDP is solving depends on discrete `gauge' symmetry in the definition of the operator basis and construction of the moment matrix. 

One can also ask why the SDP solver prefers this translation symmetric solution? The answer is clear: the Casimir energy at criticality is negative, so the periodic solution has lowest energy. In practice, this means that when studying systems at criticality, one must always assume that one is not in the thermodynamic limit. Instead, one needs to study the system at finite volume and include finite size corrections. This usually makes it  harder to compute critical exponents: one ends up having information only of half the size of the system.

\subsubsection{Discussion: transverse-field Ising SDP}
Investigating the bootstrap approach to the transverse-field Ising spin chain reveals that while convergence to the correct ground-state energy happens quickly and efficiently even for small operator bases, the convergence of the off-diagonal elements of the moment matrix $M(\B)$ is highly nontrivial and to date not well-understood. While convergence properties of the objective function are known in the literature, we are at present unaware of any mathematical results that describe in detail the convergence of off-diagonal elements of the LMI-type semidefinite program defined by \eqref{eq:bootstrapSDP}. 

Even when the energy estimate is with a few percent error, the estimates of the correlation functions can be quite inaccurate. Fundamentally, the ``correlation functions" that do not enter into the objective function are really just slack variables for the SDP; intepreting them literally risks conflating the true correlation functions of the system with optimization variables which obey only a small subset of the constraints satisfied by the system's true correlators. Only when using a physically-informed basis---that of the Jordan-Wigner fermion operators---does the method present a solid attempt at computing interesting physical quantities. However, this basis is constructed with the understanding that it describes the model in terms of its integrable variables, where the integral is the fermion parity \cite{Mbeng:2020awt}. This means that extending the present method to non-integrable models may fail if the task is accurate computation of the ground-state correlations. 

\section{The three-state Potts model} \label{sec:pots}
A natural progression in our test of the SDP's adeptness at tackling spin chain Hamiltonians is to go beyond the Ising class. A simple extension is to the highly studied $q$-state clock models \cite{Hhao2020,Meng2015}. The Ising class is the case $q=2$; in this section, we focus on the $q=3$ clock model, or equivalently the 3-state Potts model. The 3-state Potts model is an extension of the Ising model's $\mathbb{Z}_2$ symmetry to $\mathbb{Z}_3$. This is a system known to be integrable which can be decomposed into parafermion operators (analogous to the Jordan-Wigner fermions in the Ising case) using a Fradkin-Kadanoff transformation \cite{FRADKIN19801}. However, while the Ising model decomposes into free fermions, the parafermion description of the Potts model is not free in the traditional sense \cite{Fendley_2014}. 

The system's discrete lattice Hamiltonian is 
\begin{equation}\label{eq:Hamiltonian}
\hat{H}=-\sum_{i=1}^N U_i U_{i+1}^{-1}+\mu V_i+\text{h.c.}, \quad N+1=1
\end{equation}
where $N$ is the number of lattice sites and $\mu$ is a real number interpreted similarly to the transverse field in the Ising case. The operators $U$ and $V$ can be defined as matrices as
\begin{align}
U=\text{Diag}(\omega^0,\omega^1,\omega^2), \quad V=\left(\begin{array}{cc}
0 & I_{2} \\
1 & 0
\end{array}\right)
\end{align}
where $\omega=e^{\frac{2 \pi i}{3}}$ and $I_{x}$ is the $x$-dimensional identity matrix. They obey the relations $UV=\omega VU$, $U^2=U^{-1},$ $V^2=V^{-1}$ and $U^3=V^3=I_3$. Unlike in the Ising case, the matrices are not Hermitian and thus their expectation values may be complex. This will require the SDP to simultaneously solve the real and imaginary constraints. 
\begin{figure}
    \centering
    \includegraphics[scale=.45]{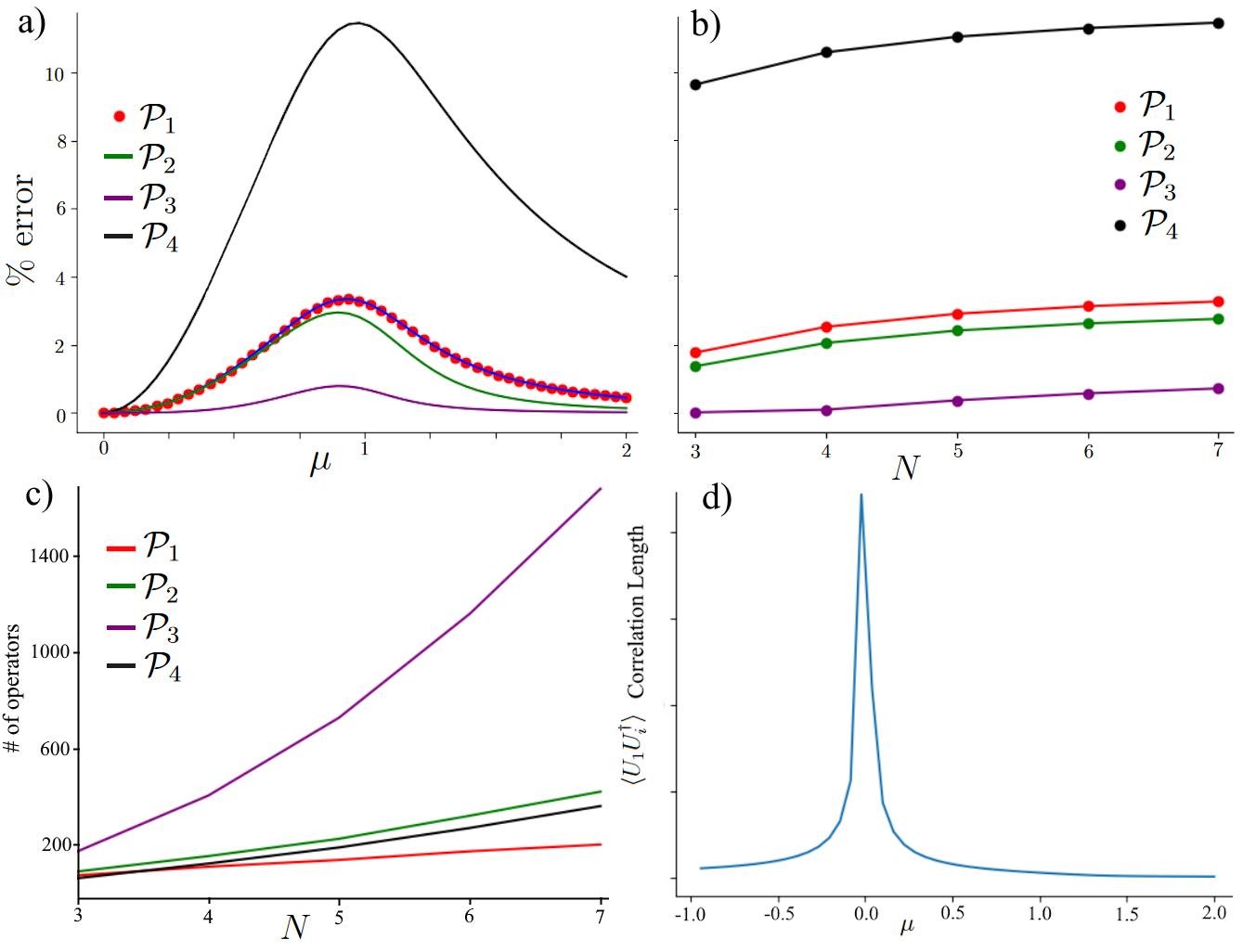}
    \caption{$a)$ Percent error of the ground state of \eqref{eq:Hamiltonian} as a function of $\mu$ for $N$=7. $b)$ Percent error of ground state energy as a function of $N$ at $\mu=1$. $c)$ The number of unique operators that make up the SDP optimization variables as a function of $N$. $\mathcal{P}_3$ has the most unique operators and thus one reasonably expects it to outperform the other bases, which it does. This logic does not generically hold as we note that $\mathcal{P}_1$ has fewer unique operators than $\mathcal{P}_4$ while outperforming in ground state estimation, as $\mathcal{P}_1 \not\subset \mathcal{P}_4$. The composite basis $\mathcal{P}_1 \cup \mathcal{P}_4$ does not improve on $\mathcal{P}_1$. $d)$ Correlation length $\chi$, e.g. $\langle U_1 U_{1+n}^\dagger\rangle\sim e^{-n/\chi},$ as a function of $\mu$. The SDP does not demonstrate spontaneous symmetry breaking at $\mu=0$; instead, the ground state density matrix it finds is a superposition of the $\mathbb{Z}_3$ degenerate ground states. Thus a phase transition is not immediately evident at $\mu=1$. This behavior is basis and $N$ independent.}
    \label{fig:percergndst}
\end{figure}

The 3-state Potts model is well-known to approach a CFT in the thermodynamic limit for $\mu\rightarrow1$ \cite{Fateev:1985mm,Ardonne2021}. The physics at $\mu=0, \infty$ is also easy to understand. At $\mu=\infty$, the terms with the $V$ dominate and one can ignore the hopping ($UU^{-1}$) terms. The ground state is a product state that diagonalizes $V$ with the eigenvalue $V=1$. The unique ground state is invariant under the ${\mathbb Z}_3$ charge generated by $Q= \prod(V_j)$. At $\mu=0$, one instead needs to diagonalize the $UU^{-1}$ operators. These are minimized if $U_j$ have all the same eigenvalue. Then $UU^{-1}$ is the identity on the ground state. There are three possible values of $U_j$, namely, the three different cube roots of unity. The 
operators $U$ are charged under the ${\mathbb Z}_3$ symmetry, so the three possible ground states spontaneously break the ${\mathbb Z}_3$. Transitions between these states are suppressed, as one needs to flip a large number of $Q$ variables to go from one vacuum to the next; infinitely many in the thermodynamic limit. The same is true for each of the three ground states. 
The phase transition in between the limits $\mu = 0,\ \mu = \infty$ can be associated to critical spontaneous ${\mathbb Z}_3$ breaking. 

To explore the application of semidefinite methods to this model, we would like to perform an analysis similar to that which was done for the Ising case. In particular, we would like to test whether we can recover conformal scaling dimensions, verify the location of the critical point, and recover the central charge. While in the Ising case the lattice operator correspondence was straightforward, here we only have a simple correspondence in the $U$ variable \cite{Mong_2014}. The process to extract the conformal data is analogous to the Ising case with slight variations. 

\begin{figure}
    \centering
    \includegraphics[scale=.5]{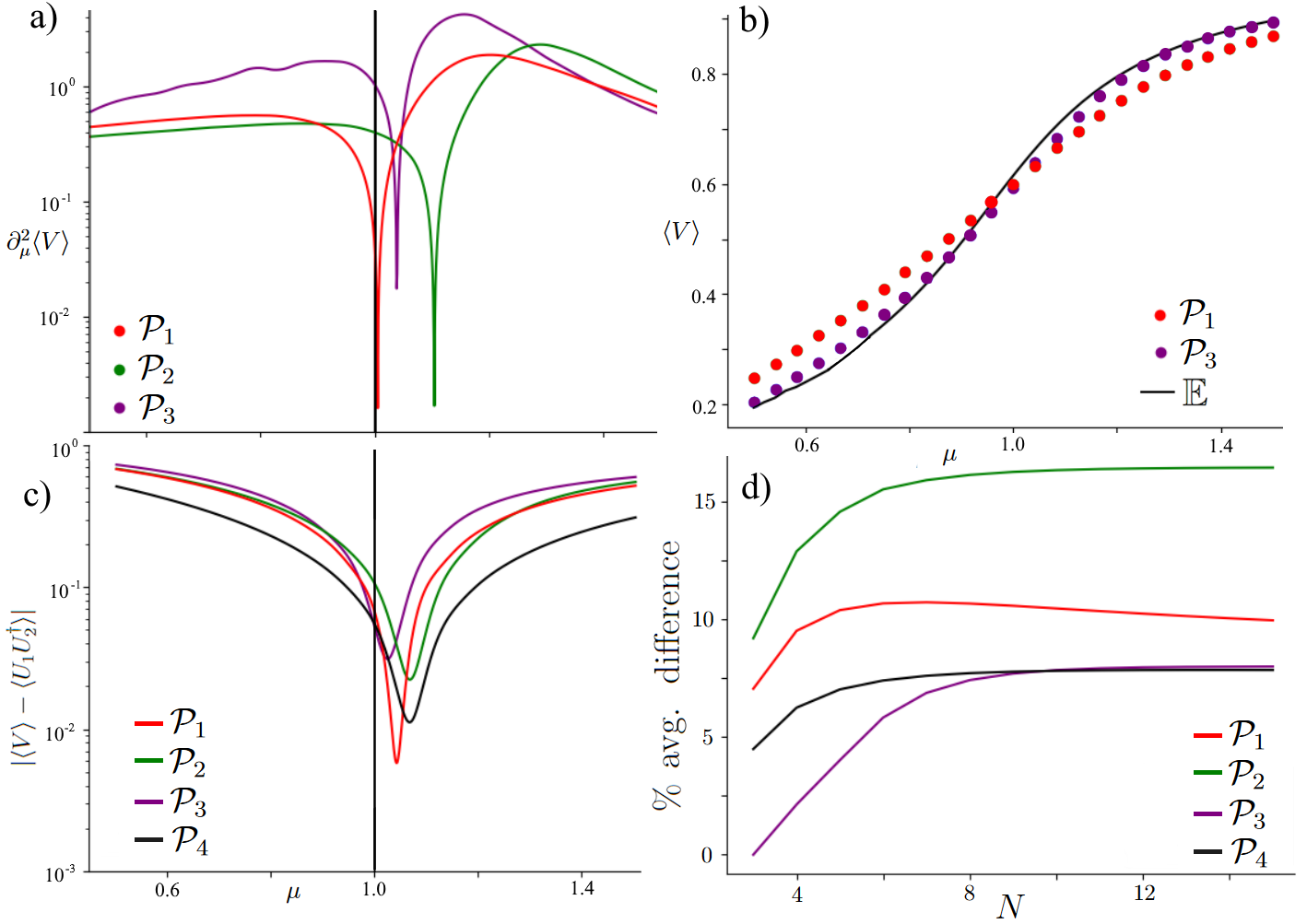}
    \caption{$a)$ $\frac{\partial^2\langle V\rangle}{\partial \mu^2}$ as a function of $\mu$ for $N=15$. $\mathcal{P}_1$ performs the best in identifying the critical point at $\mu = 1$; $\mathcal{P}_4$ is not included as it demonstrated no inflection point. $b)$ Calculated values of $\langle V \rangle$ versus $\mu$ at $N=7$ for two bases and the exactly diagonalized lattice ($\mathbb{E}$). While $\mathcal{P}_3$ is consistently closer than $\mathcal{P}_1$ to the exact lattice value, $\mathcal{P}_1$ happens to more accurately identify the divergence in $\partial_\mu^2\ex{V}$ due to idiosyncrasies in the SDP solution. $c)$ The absolute difference between $\langle V \rangle$ and $\langle U_1U_2^\dagger\rangle$ for $N=15$ as a diagnostic for the presence of the non-invertible symmetry; a signal is present at $\mu \approx 1$. $d)$ The rescaled difference $2(\langle V \rangle -\langle U_1U_2^\dagger\rangle)/(\langle V \rangle +\langle U_1U_2^\dagger\rangle)$ versus lattice size $N$ for $\mu=1$; all bases approach a plateau in $N$.}
    \label{fig:critspot}
\end{figure}

While this model is integrable, it is known to be a strongly interacting theory. 
Even though a decomposition into lattice Fermi operators was successful in the Ising case, we do not encounter the same benefits here with the respective lattice Fradkin-Kadanoff parafermion operator expansion. Furthermore, while the size of the set of all operators in the Ising case was $4^N$, in this model the complete operator set is of size $9^N$. This size mismatch is apparent given that $U$ and $V$ are not Hermitian and thus operator expectation values can be complex.
\begin{figure}
\includegraphics[scale=.4]{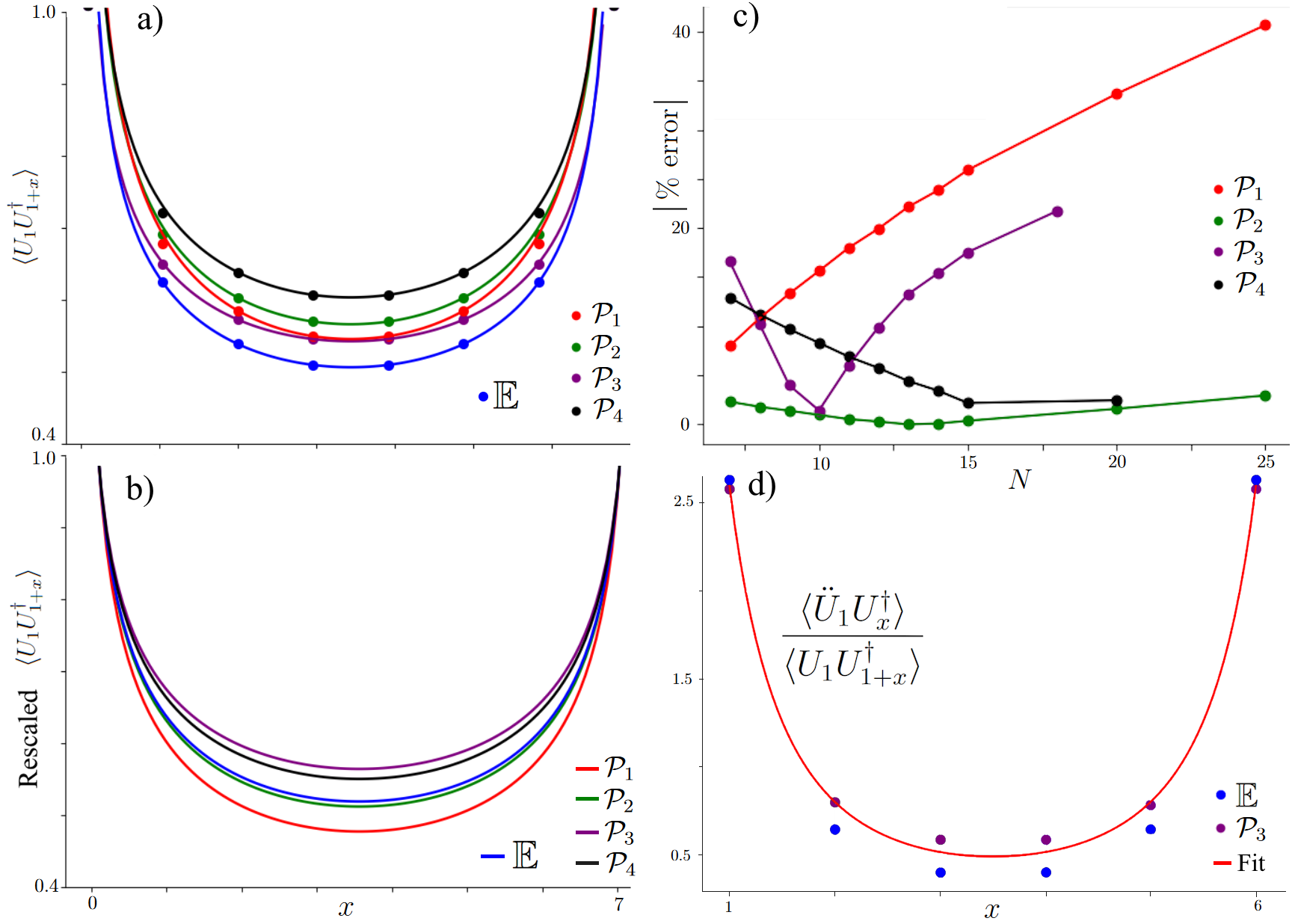}
\caption{$a)$ The correlation function $\langle U_1 U_{1+x}^\dagger\rangle$ from the SDP for $N=7$, $\mu=1$ and different bases vs exact diagonalization results  ($\mathbb{E}$); these data allow estimation of the scaling dimension $\Delta_U$. $b)$ Fits of the data in $a)$ to Eq.~\eqref{eq:fitdelta} with fixed normalization; exact correlations denoted as $\mathbb{E}$. $c)$ Percent error of the estimated scaling dimension vs the continuum scaling dimension, $\Delta_U=2/15$ as a function of lattice size $N$. The best-performing basis here, $\mathcal{P}_2$, was not best-performing for the ground state estimation. $d)$ Best basis approximation for the normalized second time derivative of the correlator.
}
\label{fig:critdim}
\end{figure}

In order to test the SDP approach we begin as before by creating a class to describe the algebra of the model. Specifically, the normal ordering of the algebra is set so all conjugate operators are brought to the left of the operator string; we also choose the ordering such that $V$ operators are brought to the left of $U$ operators (e.g., $(V_1U_1)^\dagger$, $U_1^\dagger V_1$, $V_1U_1$, etc.). This defines a total ordering on operator strings. Then, as before, an operator basis is chosen for the SDP. The problem is then solved in the dual formulation of the SDP. In this formulation, the constraints are positive semidefiniteness of the moment matrix $M$ and the conjugate relations that exist between operators. 

After exploring different operator bases, we found that an effective one includes all local operators as well as all possible two-point functions. The bases we present are as follows: $\mathcal{P}_1$ are all single site operators, $\mathcal{P}_2$ are $U_1U_{i+1}^\dagger$ and $(U_1U_{i+1}^\dagger)^\dagger$ in addition to $\mathcal{P}_1$, $\mathcal{P}_3$ is all two point operators in addition to $\mathcal{P}_1$, and $\mathcal{P}_4$ is the Fradkin-Kadanoff parafermion basis. As seen in the included figures, the parafermion lattice operators did not seem to benefit the SDP as the Jordan-Wigner fermions had in the Ising case; we imagine this is because we do not have a free fermion theory. As seen in Fig.~\ref{fig:percergndst}, the error of the ground state as a function of 
$N$ at fixed $\mu=1$ tells us that $\mathcal{P}_3$ is the basis that most closely approximates the lattice with respect to the primary objective of the SDP, the ground state energy. We also see that the SDP did not exhibit spontaneous symmetry breaking at $\mu=0$; instead, the SDP solution converged to a maximally mixed density matrix of the three degenerate $\mathbb{Z}_3$ ground states. Since these are degenerate in energy, the minimization procedure does not distinguish them and the optimizer usually finds the most symmetric solution first (where all ${\mathbb Z}_3$ charged operators vanish identically).

For the proper determination of the order parameter, which is $\langle{U}\rangle$ in a vacuum that satisfies the cluster decomposition principle
away from criticality, one needs to make sure that the degeneracy of the vacua is addressed. One way to do so is to note that $\langle{U_i U^{-1}_j}\rangle$ saturates at very long distances $|i-j|\gg 1$ to the product $\langle{U_i}\rangle \langle U^{-1}_j\rangle$, which is independent of the vacuum of choice (after all, the phase of $U$ and $U^{-1}$ cancel each other). This two point function can be used to extract $\langle U\rangle $ up to a phase (a root of unity). Here the problem is that one would need to go to very long distance near criticality to see the saturation of the correlation function.

The presence of the mixed state (maximally mixed between the vacua) in the SDP results was identified by the divergence in the naive correlation length of the two point function $\langle U_1U_{1+i}\rangle$ as $\mu\rightarrow0$ in Fig.~\ref{fig:percergndst}, indicating that the naive correlation function does not decay with distance and, even though $\langle U \rangle=0$, the two point function does not satisfy clustering.
The behavior is further verified by examining the behavior of $\langle V\rangle$ which did not approach zero and instead behaved as would be expected in the superposition state. Thus, in order to determine the critical point without \textit{a priori} knowledge, we examine the second derivative $\partial_\mu^2 \langle V \rangle$, where the phase transition should register as a discontinuity. As seen in Fig.~\ref{fig:critspot}, a feature consistently appears across bases near $\mu=1$. Similar signals consistent with the known non-invertible symmetry of the model also appear near $\mu=1$ in Fig.~\ref{fig:critspot}c . 

Having roughly identified $\mu=1$ as the critical point, we explore what conformal data we can extract. Unfortunately, extracting conformal data using the SDP, while theoretically possible, requires in this case a large basis of operators (a requirement further exacerbated by the larger algebra of the $U$, $V$ operators compared to the Ising case). Bases which give good estimates of the ground state do not necessarily give good estimations of conformal data, and the accuracies of these respective estimations may even be anti-correlated. One can recover wildly different scaling dimensions for the same exact ground state energy result depending on what constraints ones specifies in the solver (for example, conjugation relations between operators). Because the SDP can work away from criticality, the identification of critical points may be a more reasonable first application. 

In Fig.~\ref{fig:critdim} we show the calculation of the lowest-lying and simplest CFT operator, the spin field \cite{Fendley_2014}. Surprisingly, the best basis was $\mathcal{P}_2$, which had been outperformed in energy estimation. Using Eq.~\eqref{eq:centchargenorm} the speed of light $v$ was estimated from the data of Fig.~\ref{fig:critdim}d. Using the normalization presented in \cite{FAlcaraz_1987}, $v$ was estimated at $\approx 1.7$ in relative agreement to the value of 1.5 presented in the same reference. However, the central charge calculation is highly dependent on the ground state energy data. If SDP ground state data is used a central charge of $c=.44$ is found, while if exact lattice diagonalization is used then one finds $c=.812$, relatively close to the exactly known result $c = 4/5 = 0.8$. While energy estimation is relatively easy for this model, an accurate determination of the central charge remains difficult. 

From our brief excursion beyond the Ising class we see that increased complexity of the base algebra describing the system complicates the use of the SDP in extracting conformal data. Basis size alone does not describe the best basis from which to extract conformal data. As a result the lessons from the Ising case are not directly transferable to the present model.  Without \textit{a-priori} knowledge, it would be difficult to implement a system through which to correctly identify critical points and then extract accurate conformal data. 

\section{A non-integrable model}\label{sec:ANNI}
Having now addressed two integrable models, where our extra physical information informed our approach, we now turn to analyzing a non-integrable model and observing the performance of semidefinite methods as benchmarked against the exact diagonalization results. 

Following the outline of Milsted and Vidal \cite{Milsted_Vidal_2017}, we consider the axial next-nearest-neighbor Ising (ANNNI) model with Hamiltonian
\begin{equation}
    \label{eq:annni_ham}
    \hat{H}_{ANNNI} = -\sum_{j=1}^N\left[ X_{j}X_{j+1} + Z_j + \gamma X_j X_{j+2} + \gamma Z_jZ_{j+1}\right]
\end{equation}
This model was introduced and studied in \cite{Selke_1988_annni,PhysRevB.92.085139_annni,PhysRevB.92.235123_annni} and presents interesting physical behavior in the context of our previous investigation of the Ising model: at various values of $\gamma$ it describes three different emergent CFTs. Furthermore, the Jordan-Wigner description of the model is a theory of interacting fermions. We will see how the bootstrap SDP method performs both in producing the correct energies and correlation functions as obtained by exact diagonalization. 

To do this, we will use some modified, larger operator bases. In particular, we will consider the basis $\B_2$ and the combined bases $\B_F^{(i)} \equiv \B_i \cup \B_F$, where the $\B_i$ are defined as before. These bases take advantage of the performance of the fermion operators in constraining the results near the Ising point while also hopefully producing higher-fidelity representations of the correlation functions, as we saw was necessary to estimate the lattice speed of light in the Ising model. 
\begin{figure}[!h]
    \centering
    \includegraphics[width = \columnwidth]{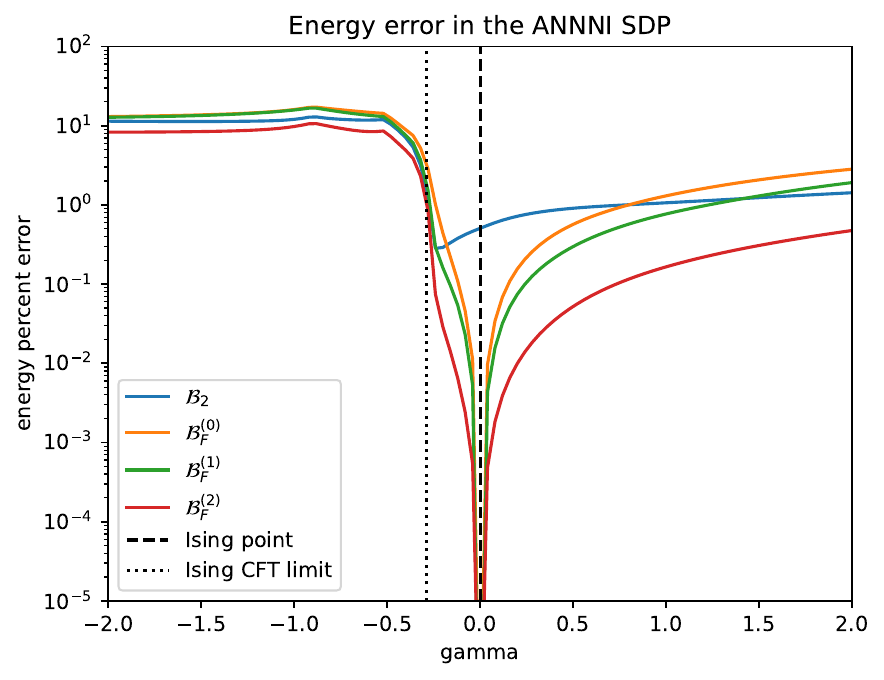}
    \caption{Energy error from SDP estimates of the ANNNI model ground state as a function of the parameter $\gamma$. The Ising point and associated free fermion theory is shown as a dashed line; the near-exact performance of the fermion basis is a clear signal of the integrability of the model. Below the Ising CFT threshold $\gamma \geq -0.285$ (dotted), the SDP error plateaus to a constant and large value.}
    \label{fig:annni_sdp_error}
\end{figure}
To begin, in Fig. \ref{fig:annni_sdp_error}, we plot the percent error of the SDP versus exact diagonalization results for the model \eqref{eq:annni_ham} at system size $N = 12$. One can clearly see the Ising point in the error curve: at the point where the system is integrable, the energy becomes heavily constrained and the error in the estimate essentially saturates at floating-point precision. A distinct decrease in the energy error is also visible moving from left to right at $\gamma \approx -0.285$, where the model begins to admit an emergent Ising CFT description. To the left of this boundary, the error plateaus to a relatively large value (and remains there for more negative values of the parameter $\gamma$). It is not clear to us why the performance of the SDP is so poor on the left side of this threshold. 

To better understand the predictions being made by the model, we can examine the results obtained for the connected $\ex{XX}$ correlation function at values of $\gamma$ in each regime of interest: we consider $\gamma = -1,-0.1,2$. 
\begin{figure*}[t]
    \centering
    \includegraphics[width = \textwidth]{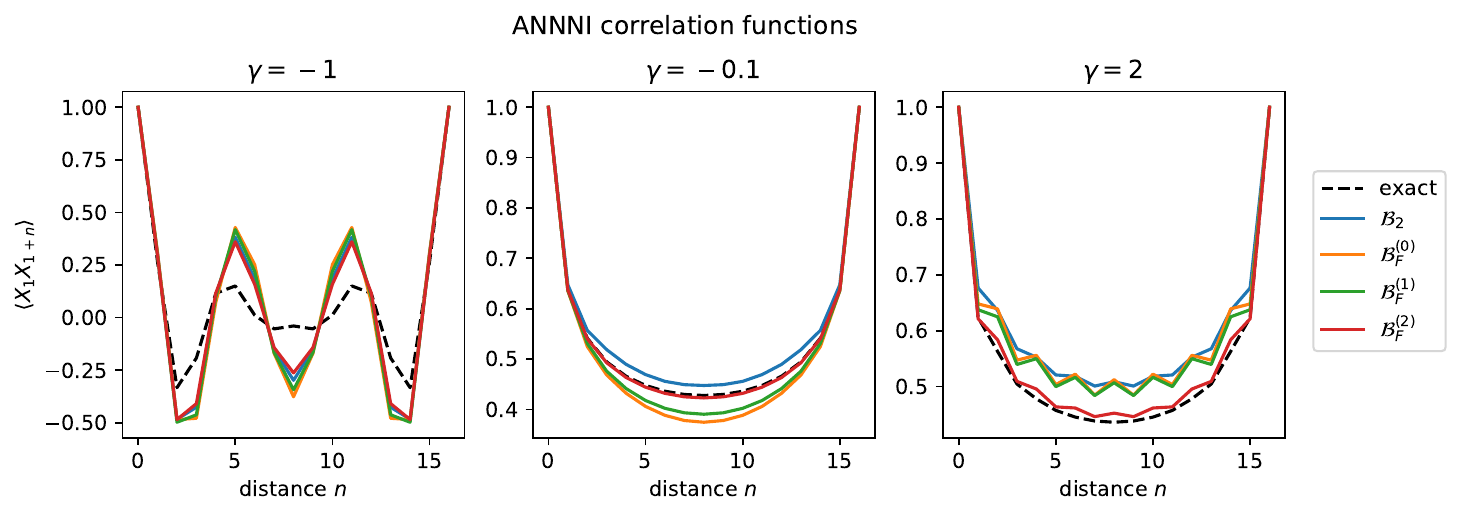}
    \caption{Comparing exact diagonalization versus SDP estimates of the correlation function $\ex{XX}$ ($N = 16$) in three regimes of the ANNNI model phase diagram: $\gamma = -1,-0.1,2$. Unsurprisingly, the largest basis does best. The SDP results are worst in the $\gamma < -0.286$ regime, consistent with the results seen in the energy estimation.}
    \label{fig:annni_corrs}
\end{figure*}
Again, while qualitative adherence to the exact values is observed, smaller bases clearly struggle to accurately reproduce the correlation functions. The agreement is worst in the regime below the Ising CFT threshold, where the energy error had plateaued to a constant and macroscopic value. The best-performing basis $\B_F^{(2)}$ contains all nearest-neighbor two-point operators in the theory as well as all single-site Jordan-Wigner fermions. While it does well nearer to the Ising point, discrepancies are clearly visible as we move away from $\gamma=0$. At $\gamma = 2$, the best-performing basis considered thus far gives an estimate of the $XX$ scaling dimension $\Delta_X \simeq 0.114$, amounting to $\sim 8\%$ error. 

We would like to be able to estimate central charges from the finite-size scaling of the energy density in the non-integrable regime of this model. Recall that doing so requires an estimation of the lattice speed of light from the correlation functions of the theory. For the ANNNI model, extracting this estimate requires using a large basis; we used the set of all one-point functions and the two-point functions $XX,YY,ZZ,XY,XZ$ (at all separations) in addition to the JW fermion operators. This is much larger than any basis considered so far in this work. 

We use this basis to solve the SDP at a decent system size and with $\gamma = 2$; from the results we extract the approximations of the correlators $\ex{X_1X_{1+n}}$ and $\ex{\ddot{X}_1X_{1+n}}$. The latter is defined by commuting with the Hamiltonian \eqref{eq:annni_ham} and gives a linear combination of 1,2 and 3-point functions. First, we regress an estimate of the scaling dimension from the standard $XX$ correlator, then we
compute the ratio in Eq. \eqref{eq:centralchargeratio_ising}. With the large basis we used, we were only able to reach $N = 12$ for this process due to memory constraints, which can be mildly remediated by moving to first-order solvers instead of MOSEK. 

The results give a scaling dimension $\Delta_X \simeq 0.12575$, an excellent result completely consistent with expectations. Regressing the speed of light gives $v_{eff} \approx 16.73$; we did not estimate this parameter by other means. Finally, we use the same basis to approximate the energy density over a range of system sizes. Combining all these estimates and using the scaling formula \eqref{eq:cardy_scaling}, we obtain an estimate of the central charge $c \simeq 0.4657$, a relatively accurate result! With more computational resources, larger bases and system sizes can be used to further decrease the error. There is no barrier in principle to estimating emergent central charges by this method. Indeed, if one were to find sufficiently well-performing bases, one could attempt to extract the three unique central charges of the phase diagram of the ANNNI model by these methods.

\section{Conclusion}\label{sec:Con}
In this work we have investigated semidefinite bootstrap approaches to directly determining correlation functions of quantum spin chains. The benefit of the bootstrap approach is  the lack of a parametric ansatz for the exact state. Under some circumstances, one can also have  polynomial scaling in the system size. Finally, there is a wide ecosystem of numerical algorithms available to use for solution. We have demonstrated procedures that can be used to estimate emergent conformal data and other physical predictions of the collective physics associated to quantum spin chains. For simple models, these procedures allowed us to estimate scaling dimensions and central charges of the emergent CFTs associated to the Ising and Potts models and the non-integrable self-dual ANNNI model. Across these models, we see excellent performance in ground energy estimation, qualitative agreement with correlation functions and, with well-chosen problem setups, numerically satisfactory results when regressing conformal data directly from the optima of the semidefinite programs. 

Semidefinite relaxations are known to provide hierarchies of approximations to exact optima of exponentially complex optimization problems. Both the physical and mathematical literature have studied in depth the convergence of the relaxed optima to the exact optima of the full theory. However, the perspective taken by these works usually does not address the role of the slack variables which, in the present case, have direct physical interpretations as measurements of observables in the putative state of the system. There is still much work to be done, both in the mathematics and in the physical applications, to understand more precisely how the optimal values of the optimization variables, not just the objective function, approach their exact values. As our results show, the convergence of these quantities is not bound by the same monotonicity as the convergence of the objective function. In a sense, if one is to be pessimistic, this is the place where results are not to be trusted. Without being able to understand (one could say also control) the correlation functions that one is extremizing over, one cannot reliably find the most important physical data. After all, the structure of correlations is more important than the precise numerical energy of the vacuum. That is, ubless one uses that number to show that other methods are very close (e.g., comparing to a variational ansatz). One  can say that this is another manifestation of the maxim that many body quantum ground state problems are computationally hard problems. Although in principle by adding more constraints one should be able to improve results systematically, if one needs to study a large fraction of all positive operators before a result is to be trusted, one  ends up with a problem with exponential complexity. After all, the number of operators on a system of $N$ qubits grows as $2^{2N}$. If these methods are to be really useful for precision data, new physics input seems to be needed to make progress. 

The bootstrap's strength lies in its generality. Essentially any quantum-mechanical system with a Hilbert space of unitary representations and an algebra of observables may be approximated by a bootstrap approach. There is nothing preventing one from considering this approach for spin systems in higher dimensions and on lattices of arbitrary geometry. All the same methods apply. As a practical application, a better idea of how correlation functions and other data converge with these methods would open up semidefinite methods to problems in quantum chemistry and molecular dynamics as well, where state-oriented variational methods are dominant.  

The difficulty of the bootstrap is in how to correctly specify the problem. The task of determining small yet strongly constraining bases of operators is related to an array of computationally hard problems in the study of spin systems, including the representability problem. There is a sense in which the theory at the lattice scale is different than the theory in the long wavelength limit; this is what the renormalization group paradigm tells us.
Understanding how to control these issues better, at least implicitly, amounts to solving this renormalization group problem.
Physical intuition may prove a deciding factor in the successful construction of approximations, though algorithmic methods to search for efficient bases may be possible, as was briefly investigated by Lawrence \cite{Lawrence:2021msm}. At least in one of our examples, we saw that we also needed non-local operators to improve the solution. Understanding this more carefully seems like a good avenues for progress.


A more systematic approach to incorporating commutator constraints could be taken by viewing the constraint $\ex{[H,\OO]}= 0$ (or any other linear, homogeneous constraint) as defining an ideal in the algebra of operators $\A$. One could construct Gr{\"o}bner bases for the quotient ring $\mathcal{A}/\{\ex{[H,\OO]}= 0\}$ and use the basis elements as the basis of operators in the SDP, thereby pre-solving some of the constraints. Algorithms exist for the construction of non-commutative Gr{\"o}bner bases \cite{Mora_1994}, but their efficiency is dubious for the general class of non-commutative algebraic structures present in interacting spin or clock-model systems. 

Beyond the choice of operator basis, correctly handling symmetries of the problem statement and automorphisms of the operator algebra could also assist in lowering the computational overhead of the problem. Invariant semidefinite programming---finding a block-diagonal decomposition of the moment matrix into irreducible representations of a symmetry---has been applied by various works e.g. \cite{Kazakov:2022xuh}, though we are unaware of research which addresses these ideas for optimization over non-commutative variables in full generality. Notice for example that in the Potts model at criticality we studied, we did not use all the information of the integrable model, which also has an underlying Temperley-Lieb algebra and a large number of additional conserved quantities. Maybe using the bootstrap ideas with this additional information will provide a good solution to that problem.

\section*{Acknowledgements}

D.B. would like to thank K. Schoutens, Y. Xin for various discussions.  D.B. and P.N.T.L research supported in part by the Department of Energy under Award No. DE- SC0019139. DB is also supported in part by the Delta ITP consortium, a program of the Netherlands Organisation for Scientific Research (NWO) funded by the Dutch Ministry of Education, Culture and Science (OCW).

\appendix

\section{Exact results for the transverse-field $XY$ model}
Here we reproduce the known solution of the transverse-field $XY$ model, which contains as a special case the transverse-field Ising model. The original solution of the Ising model was given in \cite{Pfeuty_1970} and expressions for the correlators in the general case are given in \cite{Barouch_McCoy_Dresden_1970,Barouch_McCoy_1971}. For other studies of the model, including the finite-size scaling, see \cite{Lieb_Schultz_Mattis_1961,henkel1987finite,Osborne_Nielsen_2002,koma1996finite}. The authors did not find a source which includes all these results in the exact form at finite $N$, which the present appendix aims to do. 

The model is a spin chain on $N$ sites. The operator algebra is $su(2)^{\otimes N}$ and is composed of the Pauli matrices $I_j,X_j,Y_j,Z_j$ at each site $j$. They obey the usual commutation rules at each site and commute off-site. The Hamiltonian is given by
\begin{equation*}
    \hat{H}_{X Y}=- \sum_{j=1}^N\left[\frac{(1+\gamma)}{2} X_j X_{j+1}+\frac{(1-\gamma)}{2} Y_j Y_{j+1} +h Z_j\right],
\end{equation*}
and depends on two parameters: the transverse field $h>0$ and the anisotropy $\gamma$, where $0 \leq \gamma \leq 1$. Here, we assume periodic boundary conditions, where $N+1 \equiv 1$.

The model is solved by performing a Jordan-Wigner transformation and subsequently diagonalizing the quadratic fermion Hamiltonian via a Boguliobov transformation. The result is a free fermion model for any values of $h,\gamma$. 

The transverse field Ising model is the case $\gamma = 1$. For any value of $\gamma$, at the quantum critical point $h =1$ and in the thermodyanmic limit $N \to \infty$, the model falls in the Ising universality class and is described by the Ising CFT with $c = 1/2$.

Restricting to the case of even $N$, we first define some auxiliary functions. The dispersion $\omega_\phi$ is given by
\begin{equation}
    \omega_\phi=\sqrt{(h-\cos (\phi))^2+\gamma^2 \sin ^2(\phi)}. 
\end{equation}
We define the set of ground state modes $K$ as
\begin{equation}
    K=\left\{\frac{(2 n-1) \pi}{N}, \quad n=1, \ldots, N / 2\right\}
\end{equation}
The ground state is given by the Jordan-Wigner fermion vacuum, and its energy is 
\begin{equation}
    E_0=-2 \sum_{k \in K} \omega_k
\end{equation}
The correlation functions depend on the two-particle Green's function: 
\begin{equation}
    G_N(R)=-\frac{2}{N} \sum_{k \in K} \cos (k R)\left(\frac{h-\cos (k)}{\omega_k}\right)\\-\gamma \sin (k R) \frac{\sin (k)}{\omega_k}
\end{equation}
This function is written in terms of the Jordan-Wigner fermions as
\begin{equation}
    G_N(R) = -\delta_{R0} + \ex{c^\dagger_0c^\dagger_R} + \ex{c^\dagger_Rc_0} + \ex{c^\dagger_0 c_R} - \ex{c_0c_R}
\end{equation}
The order parameter of the quantum phase transition at $h = 1$ is the magnetization $\ex{Z}$. In the limit $h \to \infty$, the system becomes completely ordered and $\ex{Z}\to 1$. In the disordered phase $h \to 0$, the magnetization $\ex{Z} \to 0$ in either (degenerate) ground state. The magnetization is given in terms of $G_N(R)$ as
\begin{equation}
    \langle Z\rangle=-G_N(0)=\frac{2}{N} \sum_{k \in K}\left(\frac{h-\cos (k)}{\omega_k}\right)
\end{equation}
For the purposes of this paper, we are concerned with the connected two point functions of each Pauli operator $X,Y,Z$. The simplest of these is given by 
\begin{equation}
    \left\langle Z_i Z_{i+R}\right\rangle^c=-G_N(R) G_N(-R)
\end{equation}
The $XX,\ YY$ connected correlators are given in terms of determinants (a result of the fermionic description). They are 
\begin{equation}
    \left\langle X_i X_{i+R}\right\rangle^c=\left|\begin{array}{cccc}
G_1 & G_0 & \cdots & G_{-R+2} \\
G_2 & G_1 & \cdots & G_{-R+3} \\
\vdots & \vdots & \ddots & \vdots \\
G_R & G_{R-1} & \cdots & G_1
\end{array}\right|
\end{equation}
and 
\begin{equation}
    \left\langle Y_i Y_{i+R}\right\rangle^c=\left|\begin{array}{cccc}
G_{-1} & G_{-2} & \cdots & G_{-R} \\
G_0 & G_{-1} & \cdots & G_{-R+1} \\
\vdots & \vdots & \ddots & \vdots \\
G_{R-2} & G_{R-3} & \cdots & G_{-1}
\end{array}\right|
\end{equation}
with $G_r \equiv G_N(R)$. The thermodynamic limit $N \to \infty$ may be obtained in a natural way by making the substitution
\begin{equation}
    \frac{2}{N} \sum_{k\in K} \longmapsto \frac{1}{\pi}\int_0^\pi dk
\end{equation}
These expressions are used to compute the exact finite-size correlation functions in the body of the paper.

\section{Computing the speed of light from the correlators}
In the continuum, with $\Delta=2h$, one has
\begin{align}
\langle \mathcal{O}(0,t)\mathcal{O}(x,0)^\dagger\rangle=\frac{1}{(x^2-v^2t^2)^\Delta}\\
\langle \dot{\mathcal{O}}(0,t)\mathcal{O}(x,0)^\dagger\rangle=\frac{2\Delta v^2 t}{(x^2-v^2t^2)^{\Delta+1}}
\\
\langle \ddot{\mathcal{O}}(0,t=0)\mathcal{O}(x,0)^\dagger\rangle=\frac{2\Delta v^2}{(x^2)^{\Delta+1}}
\end{align}
To obtain the lattice expression, one makes the substitution $x \mapsto (N/\pi)\sin(\pi x/N)$ and takes the ratio of correlation functions above to find
\begin{equation}\label{eq:centchargenorm}
\frac{\langle \ddot{\mathcal{O}}(0,0)\mathcal{O}(x,0)^\dagger\rangle}{\langle \mathcal{O}(0,0)\mathcal{O}(x,0)^\dagger\rangle}=\frac{2\pi^2v^2\Delta}{N^2\sin(\pi x /N)^2}
\end{equation}
When the lattice speed of light $v$ and scaling dimension $\Delta$ is found, one can normalize the energies such that one can use the long string expansion to calculate the central charge $c$ from e.g. the Luscher term \cite{Aharony2013,luscher2008,Cardy_1986}: 
\begin{equation}
E=TL-\frac{c}{12L}+\mathcal{O}(L^{-3})
\end{equation}
{\it Ising model---}Choosing $\mathcal{O}=X$, one has $-[H,[H,X]]=\ddot{X}$, and thus $\langle \ddot{X}_1 X_a\rangle=-\langle [H,[H,X_1]] X_a\rangle$ which we calculate as follows:
\begin{equation}
\begin{aligned}
[H,X_1]&=[-Z_1,X_1]\\
&=-2iY_1=\dot{X}_1\\
[H,\dot{X}]&=2i\left([X_1X_2,Y_1]+[X_1X_N,Y_1]+[Z_1,Y_1]\right)\\
&=-4\left(Z_1X_2+Z_1X_N-X_1\right)\\
&=\ddot{X}_1
\end{aligned}
\end{equation}
\begin{equation}
\langle \ddot{X}_1X_a\rangle= -4\left(Z_1X_2X_a+Z_1X_NX_a-X_1X_a\right)\propto \left(\frac{2\Delta \pi^2 v^2}{N^2\sin(\pi x/N)^{2(\Delta+1)}}\right)
\end{equation}
{\it Three-state Potts---}The lattice-continuum operator correspondence is most straightforward in the spin field, $U$. Thus we choose $\mathcal{O}(z,t)=U_{z\in \mathbb{Z}}(t)$, and make use of the double time derivative commutator $-[H,[H,U]]=\ddot{U}$. We set $U(0,0)=U_1$ and $U(z,0)=U_{1+z}$. Thus one can determine the time evolution as follows,
\begin{equation}
\begin{aligned}
[H,U_1]&=[V_1+V_1^\dagger,U_1]\\
&=V_1U_1(1-\omega)+V_1^\dagger U_1 (1-\omega^\dagger)\\
&=\dot{U_1}\\
[H,\dot{U_1}]&=\left([U_1U_2^\dagger+U_1^\dagger U_2,\dot{U}_1]\right.+\\ & \left.[U_1U_N^\dagger+U_1^\dagger U_N,\dot{U}_1]+[V_1,\dot{U}_1]+[V_1^\dagger,\dot{U}_1]\right)\\
&=(1-\omega^\dagger)\left(U_1^\dagger V_1^\dagger U_{2,N}^\dagger(\omega-\omega^\dagger)+V_1^\dagger U_{2,N}(\omega-1)\right.\\
&\left.+ U_1(1-\omega)+V_1U_1(1-\omega^\dagger)\right)\\
&+(1-\omega)\left( U_1^\dagger V_1 U_{2,N}^\dagger(\omega^\dagger-\omega)+V_1U_{2,N}(\omega^\dagger-1)\right.\\
&+V_1^\dagger U_1(1-\omega)+U_1(1-\omega^\dagger)\\
&=\ddot{U}_1\\
\end{aligned}
\end{equation}
\begin{equation}
\Rightarrow \langle \ddot{U}_1 U_{1+x}^\dagger\rangle\propto\left(\frac{2\Delta \pi^2 v^2}{N^2\sin(\pi x/N)^{2(\Delta+1)}}\right)
\end{equation}
\newpage
\bibliographystyle{apsrev4-1}
\bibliography{refs}

\end{document}